\documentclass[onecolumn,useAMS,usenatbib]{mn2e}

\usepackage{amsbsy}
\usepackage{amssymb}
\usepackage{amsmath}
\usepackage{graphicx}
\usepackage{natbib}
\def\be{\begin{equation}}
\def\ee{\end{equation}}
\def\bea{\begin{eqnarray}}
\def\eea{\end{eqnarray}}

\def\bear{\begin{eqnarray}}
\def\eear{\end{eqnarray}}

\def\x{\mathrm{x}}
\def\y{\mathrm{y}}
\def\n{\mathrm{n}}
\def\p{\mathrm{p}}

\def\veps{\varepsilon}

\newcommand{\pd}[2]{\frac{\partial {#1}}{\partial {#2}}}

\newcommand{\mtb}[1]{\mathbf{#1}}

\def\jnl@style{\it}
\def\aaref@jnl#1{{\jnl@style#1}}
\def\aaref@jnl#1{{\jnl@style#1}}

\def\aj{\aaref@jnl{AJ}}                   % Astronomical Journal
\def\apj{\aaref@jnl{ApJ}}                 % Astrophysical Journal
\def\apjl{\aaref@jnl{ApJ}}                % Astrophysical Journal, Letters
\def\apjs{\aaref@jnl{ApJS}}               % Astrophysical Journal, Supplement
\def\apss{\aaref@jnl{Ap\&SS}}             % Astrophysics and Space Science
\def\aap{\aaref@jnl{A\&A}}                % Astronomy and Astrophysics
\def\aapr{\aaref@jnl{A\&A~Rev.}}          % Astronomy and Astrophysics Reviews
\def\aaps{\aaref@jnl{A\&AS}}              % Astronomy and Astrophysics, Supplement
\def\mnras{\aaref@jnl{MNRAS}}             % Monthly Notices of the RAS
\def\prd{\aaref@jnl{Phys.~Rev.~D}}        % Physical Review D
\def\prl{\aaref@jnl{Phys.~Rev.~Lett.}}    % Physical Review Letters
\def\qjras{\aaref@jnl{QJRAS}}             % Quarterly Journal of the RAS
\def\skytel{\aaref@jnl{S\&T}}             % Sky and Telescope
\def\ssr{\aaref@jnl{Space~Sci.~Rev.}}     % Space Science Reviews
\def\zap{\aaref@jnl{ZAp}}                 % Zeitschrift fuer Astrophysik
\def\nat{\aaref@jnl{Nature}}              % Nature
\def\aplett{\aaref@jnl{Astrophys.~Lett.}} % Astrophysics Letters
\def\apspr{\aaref@jnl{Astrophys.~Space~Phys.~Res.}} % Astrophysics Space Physics Research
\def\physrep{\aaref@jnl{Phys.~Rep.}}      % Physics Reports
\def\physscr{\aaref@jnl{Phys.~Scr}}       % Physica Scripta

\begin{document}

\title[The dynamics of pulsar glitches]{The dynamics of pulsar glitches: Contrasting phenomenology with numerical evolutions}

\author[Sidery et al]{ T. Sidery$^1$, A. Passamonti$^2$ and N.~Andersson$^2$ 
\\ \\
$^1$ FENS, Sabanci University, Orhanli, 34956 Istanbul, Turkey\\
$^2$ School of Mathematics, University of Southampton,
Southampton SO17 1BJ, United Kingdom\\
}

%%%%%%%%%%%%%%%%%%%%%%%%%%%%%  MAKETITLE  %%%%%%%%%%%%%%%%%%%%%%%%%%%%%%%%%%%%%%
\maketitle

%%%%%%%%%%%%%%%%%%%%%%%%%%%%%  ABSTRACT  %%%%%%%%%%%%%%%%%%%%%%%%%%%%%%%%%%%%%%%
\begin{abstract}
In this paper we consider a simple two-fluid model for pulsar
glitches.  We derive the basic equations that govern the spin
evolution of the system from two-fluid hydrodynamics, accounting for
the vortex mediated mutual friction force that determines the glitch
rise. This leads to a simple ``bulk'' model that can be used to
describe the main properties of a glitch event resulting from vortex
unpinning. In order to model the long term relaxation following the
glitch our model would require additional assumptions regarding the
repinning of vortices, an issue that we only touch upon briefly.
Instead, we focus on comparing the phenomenological model to results
obtained from time-evolutions of the linearised two-fluid equations,
i.e. a ``hydrodynamic'' model for glitches. This allows us to study,
for the first time, dynamics that was ``averaged'' in the bulk model,
i.e. consider the various neutron star oscillation modes that are
excited during a glitch.  The hydro-results are of
some relevance for efforts to detect gravitational waves from
glitching pulsars, although the conclusions drawn from our rather
simple model are pessimistic as far as the detectability of these
events is concerned.
\end{abstract}

%%%%%%%%%%%%%%%%%%%%%%%%%%%%%  Keywords  %%%%%%%%%%%%%%%%%%%%%%%%%%%%%%%%%%%%%%%%%%%

%%%%%%%%%%%%%%%%%%%%%%%%%%%%%%% SEC. %%%%%%%%%%%%%%%%%%%%%%%%%%%%%%%%%%%%%%%%%%%
\section{Introduction}

Neutron stars provide excellent testbeds for extreme physics theory.
Since their core density reaches beyond anything we can produce in the
laboratory, observations of such compact remnants may provide unique
constraints on models for supranuclear physics \citep{LP04}. In order
to infer useful information from astrophysical observations we need to
construct realistic models with the power to predict the evolution of
individual systems (at least to some extent).  This is a serious
challenge for the modelling community.  Even a moderately reasonable
neutron star model should account for the presence of different exotic
states of matter.  From nuclear physics and Bardeen-Cooper-Schrieffer
(BCS) theory we expect the outer neutron star core to consist of
superfluid neutrons, superconducting protons, free electrons and
muons. Deeper into the core more exotic phases of matter, like
superfluid hyperons and/or deconfined quarks exhibiting
colour-flavour-locked superconductivity, are likely to be present.
Meanwhile, a relatively thin (1 km or so) crust surrounds the fluid
core.  In the crust, matter changes from a soup of nucleons in the
interior to an elastic nuclear lattice of heavy iron near the surface.
In the inner part of the crust (beyond neutron drip) free neutrons are
expected to be superfluid.

An important question concerns whether differences in internal
structure can be deduced from observations. This is obviously
problematic since most of the data collected for these stars originate
from the star's surface, atmosphere or magnetosphere. Having said
that, there are phenomena that involve bulk dynamics and which should
depend on the internal composition. Examples of such phenomena are the
radio pulsar glitches \citep{2000MNRAS.315..534L} and the
quasi-periodic oscillations observed in the X-rays from magnetar
flares, see for example \citet{2007Ap&SS.308..625W}. The long
relaxation time associated with glitches is seen as indirect evidence
for neutron star superfluidity \citep{AI75,Rud76,Alp81}, and recent
considerations of the crust oscillations which are thought to be the
origin of the observed magnetar oscillations suggest that these may be
affected by the presence of a crust superfluid as well \citep{AGS09}.

In this paper we discuss basic models for pulsar glitches. We focus on
the glitch event itself rather than the subsequent long term
relaxation. We consider the implications of the standard two-fluid
model from two different points of view. First of all, we derive the
simple equations that govern the bulk evolution of the system from
two-fluid hydrodynamics, accounting for the vortex mediated mutual
friction force. This leads to a basic model that can be used to
describe a glitch rise resulting from global vortex unpinning, be it
in the crust
\citep{1996ApJ...457..844L,2008ApJ...672.1103M,2008MNRAS.390..175W,2009ApJ...700.1524M}
or in the core \citep{2003PhRvL..91j1101L}\footnote{In reality, our
  model is somewhat unrealistic for both crust and core
  superfluids. In the former case we have not accounted for the crust
  elasticity, while in the latter case we are ignoring the expected
  interaction between neutron vortices and proton fluxtubes. These
  effects may have a decisive impact on glitch dynamics. Nevertheless,
  the present work is state-of-the-art in this area, and we expect to
  add the relevant features to the hydrodynamical model in future
  work. }. The results demonstrate that the model requires additional
assumptions regarding the repinning of vortices in order to model the
long-term evolution. This is as expected \citep{1984ApJ...276..325A}.
Secondly, we use time-evolution of the linearised two-fluid equations
as a ``hydrodynamic'' model for the glitch event.  This allows us to
consider dynamics that was ``averaged'' in the bulk model. In
particular, we consider the various neutron star oscillation modes
that are excited during the glitch. In principle, the obtained results
could be of relevance for efforts to detect gravitational waves from
glitching pulsars. Having said that, our estimates suggest that the
gravitational-wave signals associated with the impulsive events that
we consider here are very weak.

\section{Bulk properties}

We want to consider the simplest viable model for large pulsar
glitches.  The angular momentum of any superfluid component is
determined by the density and configuration of vortices threading the
fluid. If the vortices are fixed (pinned), there can be no angular
momentum exchange between the superfluid and other components in the
star. Assuming a scenario of catastrophic unpinning it is
straightforward to formulate a simple glitch model. One can simply
assume that the system has two components, with different moments of
inertia and spin-rates. Adding the assumption that vortex pinning
allows a lag in the rotation between the observed component (e.g. the
crust) and the other component (the interior superfluid) and that the
pinning breaks once a critical lag is reached, one arrives at a
phenomenological glitch model.  In order to effect the rapid transfer
of angular momentum that leads to the observed spin-change in the
crust, one would typically assume that the rotation lag relaxes on
some timescale. The standard model assumes that this relaxation is due
to the mutual friction acting on the superfluid vortices
\citep{ALS84,1991ApJ...380..530M,ASC06}.

While such a phenomenological model is useful, its scope is
limited. Ultimately, a detailed description will require a
hydrodynamics analysis. In particular, if we want to understand the
reason for the sudden vortex unpinning (likely due to an instability,
see \citet{2stream1,2009PhRvL.102n1101G} for recent ideas) and
possible neutron star oscillations excited by the event. Developing a
detailed hydrodynamics model is a severe challenge given the many
uncertainties in the relevant physics, but we can make some progress
by making contact between the general two-fluid hydrodynamics
framework and the phenomenological bulk dynamics description. As a
suitable starting point, we will show how the bulk model can be
obtained from hydrodynamics.  This is instructive since it provides
insight into the validity of the model, and also gives us a better
idea of the origin of the different parameters (like the global
spin-up time). Moreover, we will be able to make direct comparisons
with hydrodynamics results.

\subsection{Single fluid}

It is useful to begin by outlining the analysis of a single fluid
body.  In that case, we have a velocity field $v_{i}$ which evolves
according to the Euler equations;
\begin{equation}
  \label{eq:euleq1}
  {\cal E}_{i} = \left( \pd{}{t} + v^{j} \nabla_{j} \right) v_i + \nabla_{i} \left( \tilde{\mu} + \phi \right) = 0 \, , 
\end{equation}
where $\phi$ is the gravitational potential and $\tilde{\mu} = \mu/m$
is the chemical potential divided by the particle mass.  In addition,
we have the continuity equation
\begin{equation}
{ \partial \rho \over \partial t } + \nabla_i \left( \rho v^i \right) = 0 \, ,
\end{equation}
and the Poisson equation
\begin{equation}
\nabla^2 \phi = 4 \pi G \rho \, .
\end{equation}
Note that we use a coordinate basis to represent vector quantities
 throughout this paper.  In other words, we distinguish between co-
 and contravariant quantities, using the flat space metric $g_{ij}$ to
 relate them.  That is, we have $v_i = g_{ij} v^j$. We also make use
 of the Einstein summation convention for repeated indices.  Finally,
 we have assumed that the internal energy $E$ depends only on the
 number density $n$, i.e. that the fluid is a barotrope.  Then
\begin{equation}
\mu = {dE \over dn} \, ,
\end{equation}
and the pressure $P$ of the fluid is defined (in the usual way) by
\begin{equation}
dP =   n d \mu \, .
\end{equation}
Viscosity terms have been omitted in anticipation of applying the
solid body approximation, which we do now. Assuming uniform rotation
the fluid velocity can be written
\begin{equation}
  v_{i} = \epsilon_{ijk} \Omega^{j} x^{k} = \Omega \varpi \varphi_{i} \, ,
\end{equation}
where we take $\Omega = \Omega(t)$, $\varpi$ is the cylindrical
distance from the rotation ($z$) axis and $\varphi_{i}$ is a unit
vector in the direction of the flow.  By assuming axi-symmetry it
follows that
\begin{align}
  v^{j} \nabla_{j} v_{i} & = 0 \, ,\\
  v^{j} \nabla_{j} (\tilde{\mu} + \phi) & = 0 \, .
\end{align}
We will now derive the conservation of angular momentum and energy
from the above equations.  Contracting (\ref{eq:euleq1}) with $\rho
v_{i}$ and integrating over a fixed volume $V$ gives
\begin{equation}
  \label{eq:ChangeKE1}
  \pd{E}{t} = \int \rho v^{j} \pd{v_{j}}{t} dV = \pd{}{t} \left( \frac{1}{2} \int \rho v^{2} dV \right) = 0 \, .
\end{equation}
This shows that the kinetic energy is conserved.  Adding the
assumption that the fluid is in solid body rotation we see that
\begin{equation}
  \label{eq:velSq}
  v^{2} = \Omega_{j} \Omega^{k} \left( \delta^{j}_{k} x^{2} - x^{j} x_{k} \right) \, ,
\end{equation}
where $x^{2} = g_{ij} x^i x^j = x^{j} x_{j}$.  From this we define the
moment of inertia as
\begin{equation}
  \label{eq:inertia}
  I^{i}_{l} = \int \rho \left( \delta^{i}_{l} x^2 - x^{i} x_{l} \right) dV \, .
\end{equation}
We can use (\ref{eq:velSq}) and (\ref{eq:inertia}) to rewrite the
change in kinetic energy equation (\ref{eq:ChangeKE1}) as
\begin{equation}
  \label{eq:ChaKE1RW}
  \pd{E}{t} = \frac{1}{2} \pd{}{t} \left( I^{i}_{l} \Omega_{i} \Omega^{l} \right) = 0 \, .
\end{equation}

Let us now consider the $z$ component of the angular momentum.
Assuming that the chemical and gravitational potential only depend on
the (spherical) radial position, i.e. assuming slow rotation, then
\begin{equation}
  \epsilon_{ijk} x^{j} \nabla^{k} \left( \tilde{\mu} + \phi \right) = 0 \quad \mbox{ for } 
\quad i=z  \, .
\end{equation}
Contracting ${\cal E}_{i}$ with $\rho \epsilon_{ijk} x^{j}$ and
integrating gives
\begin{align}
  \pd{J_{i}}{t} &= \int \rho \epsilon_{ijk} x^{j} {\cal E}^{k} dV \nonumber \\
  &= \pd{}{t} \left[ \int \rho \Omega_{j} \left( \delta_{i}^{j} x^2 - x_{i} x^{j} \right) dV \right]
   = \pd{}{t} \left( I_{i}^{j} \Omega_{j} \right)
   =  0 \quad \mbox{ for }
   \quad i=z \, .
\end{align}

We see that, for cylindrical polar coordinates with $\Omega^{i} =
(0,0,\Omega)$ and $I^{z}_{z}=I$ we get the standard results
\begin{equation}
  E = \frac{1}{2} I \Omega^{2} \quad \mbox{ and } \quad
  J^{z} = I \Omega \, .
\end{equation}
Both these quantities are conserved.

%%%%%%%%%%%%%%%%%%%%%%%%%%% SubSec %%%%%%%%%%%%%%%%%%%%%%%%%%%%%%%
\subsection{Two-fluid model}

We now consider a two constituent stellar model. Our particular
interest concerns two effects, the entrainment and the mutual friction
between the components.  In order to simplify the initial analysis, we
first ignore the mutual friction.

Our formulation for multi-fluid hydrodynamics in Newtonian gravity
derives from the work by \cite{2004PhRvD..69d3001P}, see
also~\cite{2006CQGra..23.5505A}.  The analysis is based on a
variational principle, where the action is minimised by varying the
fluid flow lines.  Key in this analysis is the allowance that the
internal energy may depend on the velocity difference between the two
constituents, $w^{\n\p}_{i} = v_{i}^\n - v_{i}^\p$, such that
\begin{equation}
  d E = \pd{E}{n_\n} d n_\n + \pd{E}{n_\p} d n_\p + \pd{E}{w_{\n\p}^{2}} d w_{\n\p}^{2} = \mu_\n d n_\n + \mu_\p d n_\p + \alpha d w_{\n\p}^{2}
\label{eq:intE}  \, .
\end{equation}
Here, the constituent indices $\n$ and $\p$ denote the neutron and
``proton'' (incorporating also electrons and muons in the usual way)
components, respectively.  An important consequence is that the
conjugate momentum density for each constituent is modified so that
\begin{equation}
  p_{i}^\x = \rho_\x \left[ v_{i}^\x + \varepsilon_\x \left( v_{i}^\y - v_{i}^\x\right)\right] \, ,
\end{equation}
where $\x$ and $\y$ is either $\n$ or $\p$, with $\x \neq \y$. In this
relation the entrainment is represented by the parameter
$\veps_\x$. This is a non-dissipative effect, in the neutron star case
due to the strong interaction between neutrons and protons, which
leads to the momentum no longer being aligned with the individual
components velocity. We refer the read to
\citet{2004MNRAS.348..625P,2005NuPhA.748..675C,2005NuPhA.761..333G,2006MNRAS.368..796C}
for discussions of the role of entrainment in neutron star dynamics.
Note that,
\begin{equation}
  \varepsilon_\x \rho_\x =  \varepsilon_\y \rho_\y = 2 \alpha \, .
\end{equation}

We now have a set of Euler equations for each constituent. These
equations can be written
\begin{equation}
 {\cal E}^\x_{i} =
\left( \pd{}{t} + v_\x^{j} \nabla_{j} \right) \left( v^\x_{i} + \varepsilon_\x w^{\y\x}_{i} \right)
  + \nabla_{i} \left(\phi + \tilde{\mu}_\x \right) + \varepsilon_\x  w^{\y\x}_{j} \nabla_{i} v_\x^{j} = 0 \, .
\label{eq:Euler} 
\end{equation}
In addition, we have one continuity equation for each component and
the Poisson equation for $\phi$ is now sourced by the total mass
density $\rho = \rho_\n + \rho_\p$.  Following the analysis in the
previous section, we want to derive the equations that represent the
global conservation of energy and angular momentum. In doing this we
will, for simplicity, assume that the velocity fields of the
constituents are those of rotating solid bodies.  We also assume that
the two components rotate around the same axis, i.e. we ignore any
precessional motion.  This means that we can write
\begin{equation}
  v^{i}_\x = \Omega_\x \varpi \varphi^{i} \quad \textrm{and}\quad
  w^{i}_{\y\x} = v^{i}_\y - v^{i}_\x = \left( \Omega_\y - \Omega_\x \right) \varpi \varphi^{i} \, .
\end{equation}
As we are assuming solid body rotation, $\Omega_\x$ is not a function
of position.  Moreover, the axial symmetry of the system implies that
\begin{gather}
  v^{j}_\x \nabla_{j} v_{i}^\x = 0 \, ,\\
  v^{j}_\x \nabla_{j} \left(\phi + \tilde{\mu}_\x \right) = 0 \, .
\end{gather}
We continue to follow the method used in the single fluid case,
contract ${\cal E}^\x_{i}$ with $\rho_\x v_{i}^\x$ and integrate over
a fixed volume to find the global change in energy of each
constituent.  This leads to
\begin{equation}
\pd{E_\x}{t} = \int \rho_\x v_\x^{i} {\cal E}_{i}^\x dV =
\frac{1}{2} \int \left[ ( \rho_\x - 2 \alpha ) \pd{}{t} v_\x^{2} + 4 \alpha
v_\x^{j} \pd{v_{j}^\y}{t}\right] dV = 0 \, .
\end{equation}
We obtain the final result for the total change in energy by adding
the expressions for the two components,
\begin{equation}
  \label{eq:TotEnerg2}
  \pd{E}{t} = \pd{E_\n}{t} + \pd{E_\p}{t} = \frac{1}{2} \pd{}{t} \left\{ \int \left[ \rho_\n v_\n^{2} + \rho_\p v_\p^{2} -
4 \alpha w_{\n\p}^2 \right] dV \right\} = 0 \, .
\end{equation}
This result shows how the entrainment affects the conserved energy,
cf.  \citet{2004IJMPD..13..291C} for a detailed discussion, and agrees
perfectly with the modified ``kinetic energy'' used by
\citet{1991ApJ...380..530M}.

In the particular case of (aligned) solid-body rotation, with
$\Omega_{i}^\x$ aligned with the $z$-axis, we have
\begin{equation}
  \pd{E}{t} = \pd{}{t} \left\{ \frac{1}{2} I_\n \Omega_\n \left[ \Omega_\n 
+ \varepsilon_\n \left( \Omega_\p - \Omega_\n \right) \right] + \frac{1}{2} I_\p \Omega_\p \left[ \Omega_\p 
+ \varepsilon_\p \left( \Omega_\n - \Omega_\p \right) \right] \right\} = 0 \, .
\end{equation}
Here we have defined the constituent moment of inertia as
\begin{equation}
  \label{eq:ConIner}
  I_\x {}^{j}_{i} = \int \rho_\x \left( \delta_{i}^{j} x^{2} - x_{i} x^{j} \right) dV \, ,
\end{equation}
and $I_\x = I_\x {}^{z}_{z}$.  Similarly, we can calculate the total
change in angular momentum. To do this we note that
\begin{equation}
  \epsilon_{ijk} x^{j} v_\x^{l} \nabla^{k} v_{l}^\x =
  \epsilon_{ijk} x^{j} x^{k} \Omega_\x^{2} - \epsilon_{ijk} x^{j} \Omega^{k}_{\x} x_{l} \Omega^{l}_\x = 0  \quad \mbox{ for } \quad i=z \, ,
\end{equation}
as $\Omega_{i}^\x$ is parallel to the $z$-axis.  Contracting ${\cal
E}_{i}^\x$ with $\rho_{x} \epsilon_{ijk} x^{j}$ and integrating over
the volume $V$ we arrive at
\begin{align}
  \pd{J_{i}^\x}{t} &= \int \rho_\x \epsilon_{ijk} x^{j} {\cal E}^{k} dV \nonumber \\
  &= \int \rho_\x \epsilon_{ijk} x^{j} \left[ ( 1 - \varepsilon_\x ) \pd{v^{k}_\x}{t} +
\varepsilon_\x \pd{v_\y^{k}}{t} \right] dV = 0  \quad \mbox{ for } \quad i=z \, .
\end{align}
This can be rewritten as
\begin{equation}
  \label{eq:ConstitMomenta}
  \pd{J_{i}^\x}{t} = \pd{}{t} \left\{ I_\x {}^{j}_{i} \left[ \Omega_{j}^\x + \varepsilon_\x \left( \Omega_{j}^\y 
- \Omega_{j}^\x \right) \right] \right\} = 0 \quad \mbox{ for } \quad i=z \, ,
\end{equation}
from which it follows that the total change in angular momentum is
given by
\begin{equation}
  \label{eq:TotMomChange}
  \pd{J_{z}}{t} = \pd{}{t} \left( I_\n \Omega_\n + I_\p \Omega_\p
  \right) = 0 \, .
\end{equation}
Hence, the total angular momentum is conserved. This is obviously not
surprising. The non-trivial result concerns how the entrainment
affects the evolution of the individual components. This will be
important later.

%%%%%%%%%%%%%%%%%%%%%%%%%%%% SEC %%%%%%%%%%%%%%%%%%%%%%%%%%

\section{A simple spin-down model}\label{SpinDn}

We are not yet in a position where we can model glitches. To do this,
even in the most basic fashion, we need to account for the coupling
due to mutual friction. However, before we discuss that problem it is
interesting to consider how the conservation equations that we have
obtained can be used to model the rotational evolution of the system.
The main purpose of doing this is to understand how the entrainment
enters the problem.  As we will see, the result can be quite
surprising.

Equation~(\ref{eq:ConstitMomenta}) shows that the angular momentum of
each constituent is conserved.  In a neutron star we would expect the
protons to be locked to the magnetic field and the crust.  Hence, they
should be spun down due to a magnetic torque
$\dot{J}_{i}^\mathrm{em}$.  We can include this torque in the
equations by breaking the conservation of the proton angular momentum,
so that
\begin{equation}
  \dot{J}_{i}^\p = - \dot{J}_{i}^\mathrm{em} \, ,
\end{equation}
(from now on we will often represent time derivatives by dots). To be
specific, we assume that the magnetic torque is related to the angular
velocity of the protons by
\begin{equation}
  \dot{J}_\mathrm{em} = {\cal A} I_\p \Omega_\p^{3} \, .
\end{equation}
This is in accord with the standard magnetic dipole model.  We will
also assume that the constituent moments of inertia are constant.
Noting that $I_\n \varepsilon_\n = I_\p \varepsilon_\p$ and defining
$\varepsilon = \varepsilon_\p$ equation (\ref{eq:ConstitMomenta})
leads to
\begin{align}
  \label{eq:angMomSimp1}
   I_\n \dot{\Omega}_\n + I_\p \varepsilon \left( \dot{\Omega}_\p - \dot{\Omega}_\n \right) &= 0 \, ,\\
  \label{eq:angMomSimp2}
   I_\p \dot{\Omega}_\p + I_\p \varepsilon \left( \dot{\Omega}_\n - \dot{\Omega}_\p \right) &= - {\cal A} I_\p \Omega_\p^{3} \, .
\end{align}
Defining $\tilde{I} = I_\p/I_\n$ equation (\ref{eq:angMomSimp1}) gives
\begin{equation}
  \label{eq:DotOmegaNsimp}
  \dot{\Omega}_\n = - \frac{\varepsilon  \tilde{I}}{1 - \varepsilon \tilde{I}} \dot{\Omega}_\p \, .
\end{equation}
Substituting this into equation (\ref{eq:angMomSimp2}) we get
\begin{equation}
  \label{eq:TimeOmegaDot1}
 \bar{I} \dot{\Omega}_\p \equiv \left( 1 - \frac{\varepsilon}{1 - \varepsilon \tilde{I}} \right) \dot{\Omega}_\p = - {\cal A} \Omega_\p^{3} \, .
\end{equation}
This is a separable equation so we can integrate to get
\begin{equation}
  \int^{\Omega_\p}_{\Omega_{0}} \frac{d \Omega_\p}{\Omega_\p^{3}} = - \frac{{\cal A}}{\bar{I}} \int^{t}_{t_{0}} dt \, ,
\end{equation}
where $\Omega_{0}$ is $\Omega_\p$ at time $t_{0}$. Setting $t_{0} = 0$
we find the solution
\begin{equation}
  \Omega_\p = \Omega_{0} \left( 1 + \frac{2 {\cal A} \Omega_{0}^{2} t}{\bar{I}} \right)^{-1/2} \, .
\end{equation}
As we expect the evolution to be slow, the second term in the bracket
is small and we can expand to get
\begin{equation}
  \label{eq:OmegaPImBored}
  \Omega_\p \approx \Omega_{0} \left( 1 - \frac{ {\cal A} \Omega_{0}^{2} t}{\bar{I}} \right) \, .
\end{equation}
From this we can find the characteristic evolution timescale $\tau$
for the crust (the protons).  Ignoring entrainment we have
\begin{equation}
  \label{eq:TimeOmegaDot2}
  \tau = \frac{1}{{\cal A} \Omega_{0}^{2}} \, .
\end{equation}
${\cal A}$ can be calculated from the standard magnetic dipole model
\citep{1983bhwd.book.....S}.  Modifying the result so that the torque
acts on the proton fluid rather than the whole star we find
\begin{equation}
  {\cal A} = \frac{B_\p^{2} R^{6} \sin^{2} \theta}{6 c^{3} I_\p} \, ,
\end{equation}
where $B_\p$ is the strength of the magnetic dipole with axis at an
angle $\theta$ to the rotation axis, $R$ is the radius of the star and
$c$ is the speed of light.  As $I_\p \ll I_\n \approx I$ (typically),
we can use
\begin{equation}
  I_\p \approx \frac{I_\p}{I_\n} I \approx \frac{I_\p}{I_\n} \frac{2 M R^{2}}{5} \, ,
\end{equation}
where $M$ is the mass of the star.  Using typical parameters $B =
10^{12} \textrm{G}$, $R = 10^{6}\textrm{cm}$, $I_\p/I_\n = 0.05$,
$\Omega_{0} = 2 \pi / 0.1 \phantom{.} \textrm{s}^{-1}$ and $M =1.4
M_\odot$ and setting $\theta=\pi / 2$ we find $\tau \approx 7 \times
10^4$~years.

Let us now consider the evolution of the, unseen, neutron component.
From equation (\ref{eq:DotOmegaNsimp}) we have
\begin{equation}
  \Omega_\n - \Omega_\n^0 = - \frac{\varepsilon \tilde{I}}{1 - \varepsilon \tilde{I}} \left( \Omega_\p - \Omega_{0} \right) \, ,
\end{equation}
where $\Omega_{\n}^0 = \Omega_\n(t=0)$.
Substituting equation (\ref{eq:OmegaPImBored}) into this we get
\begin{equation}
  \label{eq:OmegaNEvolveSimp}
  \Omega_\n \approx \Omega_{\n}^0 + \frac{\varepsilon \tilde{I}}{1 - \varepsilon \tilde{I}} \frac{ \Omega_{0} t}{\tau} \, .
\end{equation}
This relation allows us to estimate how long it takes for a rotational
lag to develop between the two constituents.  Assuming that the two
components rotate together at time $t = 0$ then $\Omega_{\n}^0 =
\Omega_{0}$.  The rotational lag then evolves according to
\begin{align}
  \Delta \Omega = \Omega_\n - \Omega_\p &= \frac{\varepsilon
  \tilde{I}}{1 - \varepsilon \tilde{I}} \frac{ \Omega_{0} t}{\tau} +
  \frac{ \Omega_{0} t}{\tau} \nonumber \\ &= \left( 1 +
  \frac{\varepsilon \tilde{I}}{1 - \varepsilon \tilde{I}} \right)
  \frac{ \Omega_{0} t}{\tau} = \frac{1}{1 - \varepsilon \tilde{I}}
  \frac{ \Omega_{0} t}{\tau} \, ,
\end{align}
or
\begin{equation}
  \label{eq:timescaleLag}
  \frac{\Delta \Omega}{\Omega_\p} \approx \frac{1}{1 - \varepsilon \tilde{I}} \frac{t}{\tau} \, .
\end{equation}
It has been argued \citep{2000MNRAS.315..534L} that a rotational lag
 of ${\Delta \Omega}/{\Omega_\p}\approx 10^{-4}$ is needed in order to
 ``explain'' Vela sized glitches.  From (\ref{eq:timescaleLag}) we see
\begin{equation}
  \frac{t}{\tau} \approx 10^{-4} \to t \sim 7 \phantom{.} \textrm{years} \, .
\end{equation}
Hence, this simple model is consistent with large glitches occuring
once in a few years in a typical young pulsar.

Finally, let us consider the entrainment coupling in more detail.  In
general, the evolution of the rotation of the proton and neutron
fluids is given by (\ref{eq:OmegaPImBored}) and
(\ref{eq:OmegaNEvolveSimp}), respectively.  If we focus on a
sufficiently short evolutionary timescale, then we can assume that
$\dot{J}_\mathrm{em}$ is approximately constant.  From
equations~(\ref{eq:angMomSimp1})-(\ref{eq:DotOmegaNsimp}) we find
\begin{equation}
  \dot{\Omega}_\p = - \frac{1}{I_\p} \left( 1 - \frac{\varepsilon
  I_\n}{I_\n - \varepsilon I_\p} \right)^{-1} \dot{J}_\mathrm{em} \, .
\end{equation}
This is an interesting, and perhaps surprising result.  It appears
that, even though a spin down torque acts on the protons their
rotational velocity may increase. This happens when
\begin{equation}
  \frac{I_\n}{I_\n + I_\p} < \varepsilon < \frac{I_\n}{I_\p} \, .
\end{equation}
Would this happen for realistic parameter values?  Setting
$\varepsilon = 0.05$ and $I_\p/I_\n = 0.1$ , typical values in the
outer neutron star core \citep{2006NuPhA.773..263C}, we find that the
crust should spin down. However, if we consider the neutron fluid we
find the condition for spin up is
\begin{equation}
  \label{eq:FasterLag}
  0 < \varepsilon < \frac{I_\n}{I_\n + I_\p} \, .
\end{equation}
For the expected values of $\varepsilon$ and $I_\p/I_\n$ we see that
the neutron fluid spins up. This is somewhat
counter-intuitive, but does not violate any fundamental principles. In fact, it is easy to see how this effect can result from 
an exchange of angular momentum in a coupled system. Unfortunately, since we do not observe the neutron
component directly, the result does not help us constrain the
entrainment parameter.  It is just an  example of the
drastic effect that entrainment can have on the evolution of a system.

%%%%%%%%%%%%%%%%%%%%%%%%%%%% SEC %%%%%%%%%%%%%%%%%%%%%%%%%%
\section{Modelling a Glitch}

The two-component model that we have described cannot be used to model
glitch events unless we add more physics to it. Key to the problem is
the motion of the superfluid vortices.  There are two aspects to this
problem; We need to account for the friction that arises due to the
presence of the vortices and the associated dissipative coupling
between the two fluids in the model
\citep{ALS84,1991ApJ...380..530M,ASC06}. We should also consider the
interaction between the vortices and the nuclei in the neutron star
crust, the potential pinning of the vortices to the lattice
\citep{2006PhLB..640...74D,2008NuPhA.811..378A} and the extent to
which the vortices exhibit creep in the presence of a rotational lag
\citep{AI75,1984ApJ...276..325A,1989ApJ...346..823A,1993ApJ...403..285L,1993ApJ...409..345A}.

In the following, we will focus on the role of the mutual friction and
the glitch event itself. The vortex pinning will be dealt with in a
very simplistic fashion.  We will simply assume that the vortices are
either perfectly pinned or completely unpinned. In such a model, a
glitch would proceed as follows.  Assume that the superfluid vortices
form a uniform, straight, array aligned with the rotation axis (that
is, we are not considering potential vortex tangles and turbulence
\citep{2007MNRAS.381..747A}, which would make the problem quite a lot
harder since the model would have to contain local information
\citep{2005ApJ...635.1224P}).  In this case vortex pinning simply
fixes the number of vortices per unit area.  This in turn fixes the
neutron fluids angular momentum, so the superfluid component rotates
at a constant rate.  If we assume that the charged fluid is locked to
the crust via magnetic effects then the vortices will be rotating with
the charged fluid component.  As the crust spins down due to the
electromagnetic torque, a velocity difference will build up between
the two constituents.  This will lead to an increasing Magnus force
acting on the vortices.  Eventually, when some critical lag, $\Delta
\Omega_\mathrm{c}$, is reached this force will be strong enough to
overcome the nuclear pinning and the vortices are suddenly free to
move. At this point the vortex mutual friction becomes relevant and
serves to transfer angular momentum between the two components.  This
becomes the mechanism by which the two components couple and the lag
decays.  The crust spins up leading to the observed glitch jump.  If
the system relaxes completely, the end state should be such that the
two components rotate at the same rate.  The glitch event itself is
relatively sudden. The best resolved event to date is the so-called
Christmas glitch in the Vela pulsar, where the glitch rise time was
shorter than a few tens of seconds~\citep{2002ApJ...564L..85D}.  In
other words, the angular momentum is transfered to the crust in less
than a few hundred rotation periods.  On a longer timescale one would
expect the vortices to repin. After all, in the relaxed state the
Magnus force is absent (or at least very small).  The repinning should
determine the long-term relaxation of the glitch, i.e. the spin
evolution on timescales longer than tens of seconds. In order to model
this phase one would likely need to account for vortex creep.
Eventually, the system will reach a state where the rotational lag
increases, and the pulsar may glitch again.

We focus on the glitch event itself, i.e. the short term evolution
following global vortex unpinning. During this phase one would expect
the main dynamics to be determined by the mutual friction
force. Hence, we do not have to consider either the electromagnetic
torque or the vortex creep. These are, of course, important in a
complete glitch model, c.f. \citet{2002MNRAS.333..613L} for an
interesting discussion of the relaxation phase, but since they require
dynamics on rather different timescales it is natural to (at least
initially) consider the different phases separately. Our main aim is
to compare a simple ``global'' glitch model to a numerical evolution
that takes into account the actual hydrodynamics.

%%%%%%%%%%%%%%%%%%%%%%%%%%%% SUBSEC %%%%%%%%%%%%%%%%%%%%%%%%%%
\subsection{The evolution equations}

To model a glitch event we need to account for the mutual friction
force.  As discussed by \citet{ASC06} this means that we consider the
evolution equations
\begin{equation}
  {\cal E}^\x_{i} =
   \left( \pd{}{t} + v_\x^{j} \nabla_{j} \right) \left( v^\x_{i} + \varepsilon_\x w^{\y\x}_{i} \right)
    + \nabla_{i} \left(\phi + \tilde{\mu}_\x \right) + \varepsilon_\x w^{\y\x}_j \nabla_i v_\x^j
  = \frac{\rho_\n}{\rho_\x} \mathcal{B} \left| \omega_\n \right|  w^{\y\x}_{i} 
  - \frac{\rho_\n}{\rho_\x} \mathcal{B}' \epsilon_{ijk} \omega_\n^{j} w_{\y\x}^{k} \label{eq:EulMF} \, ,
\end{equation}
where $\mathcal{B}$ and $\mathcal{B}'$ are the mutual friction
parameters~\citep{ASC06}. We have assumed that the vortex array remains straight
and that the rotational lag remains orthogonal to the rotation axis.
This is the simplest assumption. If we were to relax it, we would have
to consider possible precession of the system and the mutual friction
force would be more complicated. In the case that we are considering,
we have
\begin{equation}
  \omega^\n_{i} = 2 \Omega_{i}^\n + \varepsilon_\n \left( 2 \Omega_{i}^\p - 2 \Omega_{i}^\n \right) \, .
\end{equation}

Generalising the prescription from the previous section, we can find
the energy equation for each constituent. This leads to
\begin{align}
\pd{E_\x}{t} &= \int \rho_\x v_\x^{j} {\cal E}_{j}^\x dV = \int
 \rho_\x v_\x^{j} \left[ \mathcal{B} \frac{\rho_\n}{\rho_\x} \left|
 \omega_\n \right| w^{\y\x}_{j} - \mathcal{B}' \frac{\rho_\n}{\rho_\x}
 \epsilon_{jkl} \omega_\n^{k} w_{\y\x}^{l} \right] dV \, .
\end{align}
The second term in the integral will vanish as $v_{i}^\x$ is parallel
to $w_{i}^{\y\x}$.  Written in terms of the moments of inertia $I_\x
{}^{j}_{i}$, cf. (\ref{eq:ConIner}), the total
change in energy is given by
\begin{equation}
  \pd{E}{t} = \pd{}{t} \left\{ \frac{1}{2} I_\n \Omega_\n \left[ \Omega_\n 
+ \varepsilon_\n \left( \Omega_\p - \Omega_\n \right) \right] + \frac{1}{2} I_\p \Omega_\p
\left[ \Omega_\p + \varepsilon_\p \left( \Omega_\n - \Omega_\p \right) \right] \right\} 
= - \mathcal{B} \left| \omega_\n  \right| ( \Omega_\p - \Omega_\n )^{2} I_\n < 0 \, .
\end{equation}
That is, the mutual friction leads to a loss of kinetic energy (as
expected).  The equilibrium (minimum energy) state is reached when the
two fluids are rotating together ($\Omega_\p = \Omega_\n$).

We can also calculate the global change in angular momentum.  Focusing
on the $z$-component of the angular momentum we find
\begin{equation}\label{eq:AngMom}
  \pd{J^\x_{i}}{t} = \int \rho_\x \epsilon_{ijk} x^{j} {\cal E}^k dV =
    \int \epsilon_{ijk} x^{j} \left[ \mathcal{B} \rho_\n \left|
    \omega_\n \right| w_{\y\x}^{k} - \mathcal{B}' \rho_\n
    \epsilon^{klm} \omega^\n_{l} w^{\y\x}_{m} \right] dV \, .
\end{equation}
Noting that
\begin{equation}
  \epsilon_{ijk} x^{j} \epsilon^{klm} \omega^\n_{l} w^{\y\x}_{m} =
  x_\x^{j} \omega^\n_{i} w^{\y\x}_{j} - x^{j} \omega^\n_{j}
  w^{\y\x}_{i} = 0 \quad \textrm{for} \quad i = z \, ,
\end{equation}
and
\begin{equation}
  \epsilon_{ijk} x^{j} w_{\y\x}^{k} = \epsilon_{ijk} x^{j}
  \epsilon^{klm} \Omega^{\y\x}_{l} x_{m} = \Omega^{\y\x}_{i} x^{j}
  x_{j} - \Omega^{\y\x}_{j} x^{j} x_{i} \, ,
\end{equation}
we can write the change in angular momentum in terms of the
constituent moments of inertia to get
\begin{equation}
  \label{eq:SepMomConst}
  \pd{J^\x_{i}}{t} = \mathcal{B} \left| \omega_\n \right| \left(
  \Omega^\y_{j} - \Omega^\x_{j} \right) I_\n {}_{i}^{j}  \, .
\end{equation}
The total angular momentum is given by
\begin{equation}
  \label{eq:TotMomTwo}
  \pd{J_{i}}{t} = \pd{J^\n_{i}}{t} + \pd{J^\p_{i}}{t} = 0 \, .
\end{equation}
Hence, the angular momentum is conserved.  This is as expected since
no external torques have been accounted for. It is worth noting that
the $\mathcal{B}'$ coefficient does not feature in the final
equations.

%%%%%%%%%%%%%%%%%%%%%%%%%%%% SUBSEC %%%%%%%%%%%%%%%%%%%%%%%%%%
\subsection{An explicit solution}

We will now solve the global evolution equations for the system. To do
this, it is useful to rewrite them in terms of the rotational lag and
a quantity directly related to the total angular momentum.  The
conservation of angular momentum makes the latter variable trivial to
deal with, while the lag is a key (more or less directly observable)
quantity.

From (\ref{eq:TotMomChange}), (\ref{eq:TotMomTwo}) and assuming that
the moment of inertia of each constituent remains constant, we define
$\mathcal{V}$ from the angular momentum such that
\begin{equation}
  \label{eq:DefnV}
   I_\n \dot{\Omega}_\n + I_\p \dot{\Omega}_\p = I \dot{\mathcal{V}} = 0 \, .
\end{equation}
Then it is obvious that $\mathcal{V}$ remains constant during the
evolution.  From (\ref{eq:TotMomChange}) and (\ref{eq:SepMomConst}) we
next find that the rate of change of the lag $ {\cal W} = \Omega_\n -
\Omega_\p$ is given by
\begin{equation}
  \label{eq:ChangeInLag}
 (1-\bar{\varepsilon})\dot{{\cal W}} = \dot{\Omega}_\n -
 \dot{\Omega}_\p = - 2 (\Omega_\n - \varepsilon_\n W) \mathcal{B}
 \left( 1 + \frac{I_\n}{I_\p} \right) {\cal W} \, ,
\end{equation}
where we have defined
$\bar{\varepsilon}=\varepsilon_\n+\varepsilon_\p$.  From
(\ref{eq:DefnV}) we can rewrite $\mathcal{V}$ as
\begin{equation}
  \label{eq:DefVconstant}
  \mathcal{V} = \frac{I_\n}{I} \Omega_\n + \frac{I_\p}{I} \Omega_\p =
    \frac{I_\n}{I} \Omega_\n + \frac{I_\p}{I} \left( \Omega_\n - {\cal
    W} \right) = \Omega_\n - \frac{I_\p}{I} {\cal W} \, .
\end{equation}
Substituting this into (\ref{eq:ChangeInLag}) gives
\begin{equation}
  \label{eq:ChangeInLAg}
  (1-\bar{\varepsilon}) \dot{{\cal W}} = - 2 \mathcal{B} \left[
  \mathcal{V} + \left( \frac{I_\p}{I} - \varepsilon_\n \right) {\cal
  W} \right] \frac{I}{I_\p} {\cal W} \, .
\end{equation}
This is the stage at which the change of variables helps us.  Because
$\mathcal{V}$ is constant, equation~(\ref{eq:ChangeInLAg}) is
separable. Straightforward integration, assuming that the glitch
occurs at time $t=0$ and defining ${\cal W}_{0} = {\cal W}(0)$, leads
to
\begin{equation}
\mathcal{W} = \mathcal{V} \left[ { \mathcal{V} \over \mathcal{W}_0}
e^{t/\tau} + \left( \frac{I_\p}{I} - \varepsilon_\n \right) \left(
e^{t/\tau} -1 \right) \right]^{-1} \, .
\label{eq:solutionW}
\end{equation}
The spin-up time $\tau$ is given by
\begin{equation}
\tau = { (1-\bar{\varepsilon}) I_\p \over 2\mathcal{B} I \mathcal{V}} \, .
\end{equation}
For practical purposes it is better to express $\mathcal{V}$ in terms
of the initial conditions.  Defining the initial rotation of the
protons as $\Omega_{0}$ we easily find $\mathcal{V}$ at time $t = 0$.
From (\ref{eq:DefnV}) it follows that
\begin{equation}
  \mathcal{V} = \frac{1}{I} \left( I_\n \Omega_\n + I_\p \Omega_\p
  \right) = \frac{I_\n}{I} \left( \mathcal{V} + \frac{I_\p}{I} {\cal
  W}_{0} \right) + \frac{I_\p}{I} \Omega_{0} \, .
\end{equation}
This rearranges to give
\begin{equation}
  \mathcal{V} = \frac{I_\n}{I} {\cal W}_{0} + \Omega_{0} \approx
  \Omega_0 \, ,
\end{equation}
which should hold since $\mathcal{W}_0 \ll 1$.  We then arrive at the
final result
\begin{equation}
\mathcal{W} \approx \Omega_0 \left[ {{\Omega_0} \over \mathcal{W}_0}
e^{t/\tau} + \left( \frac{I_\p}{I} - \varepsilon_\n \right) \left(
e^{t/\tau} -1 \right) \right]^{-1} \approx \mathcal{W}_0 e^{-t/\tau} \, ,
\end{equation}
where
\begin{equation}
\tau \approx { (1-\bar{\varepsilon}) I_\p \over 2\mathcal{B} I \Omega_0} \, .
\label{relax}
\end{equation}
It is easy to show that the observed component (the protons) evolves
according to
\begin{equation}
\Omega_\p \approx \Omega_0 + { I_\n \over I} \mathcal{W}_0 \left( 1 -
e^{-t/\tau} \right) \, .
\end{equation}

%%%%%%%%%%%%%%%%%%%%%%%%%%%% SUBSEC %%%%%%%%%%%%%%%%%%%%%%%%%%
\subsection{Matching Observational Data}\label{sec:4.3}

The model we have described is obviously quite simplistic.  Most
importantly, the assumption of constant parameters (which allowed us
to carry out the integration over the body of the star in the first
place) is quite unrealistic.  Having said that, it would not be
surprising if the final model were to retain some of the bulk dynamics
of the more complex system. Of course, the various quantities in, for
example, \eqref{relax} must be take to represent ``body averages'' in
some sense.  Moreover, the model is only relevant on the relatively
short timescale of the glitch jump itself.  In order to describe the
subsequent long-term evolution we would need to include both the the
magnetic spin-down torque and the repinning of the vortex lines to the
crust. The latter could possibly be accounted for by ``switching off''
the mutual friction ``gradually''.  That is, one could simply take
$\mathcal{B}$ to be time-dependent, reflecting the amount of vortex
pinning or the nature of the vortex creep.  This idea has been
considered by \citet{TLS_thesis}, and we think it would be interesting
to develop it further.

Despite these caveats, it is interesting to consider how observations
may constrain the various parameters. Let us consider the scenario
where the observed component represents a small fraction of the total
moment of inertia. This would be the case for a typical neutron star
crust coupled to a large superfluid reservoir in the core, when we may
have $I_\p/I \sim 10^{-2}$. Then the observed glitch jump would be
\begin{equation}
{ \Omega_\p - \Omega_0 \over \Omega_0 } \approx { I_\n \over I} {
\mathcal{W}_0 \over \Omega_0} \approx { \mathcal{W}_0 \over \Omega_0} \, .
\end{equation}
That is, $\mathcal{W}_0$ would correspond (more or less directly) to
the observed glitch size. At the same time, the available constraint
on the glitch rise time can be compared to the spin-up time of the
model. Let us, for simplicity, impose the constraint that the glitch
happens in less than 100 rotations. Then we need
\begin{equation}
\tau \Omega_0 \approx { (1-\bar{\varepsilon}) I_\p \over 2\mathcal{B}
I} < 10^2 \, .
\end{equation}
We can rewrite the entrainment factor in terms of the effective proton
mass in the usual way \citep{2002A&A...381..178P}.  Then
\begin{equation}
\varepsilon_\p = 1 - {m_\p^* \over m_\p} \qquad \longrightarrow \qquad
1 - \bar{\varepsilon} \approx {m_\p^* \over m_\p} \, ,
\end{equation}
and we need
\begin{equation}
\tau \Omega_0 \approx {m_\p^* \over m_\p} { I_\p \over 2\mathcal{B} I}
< 10^2 \, ,
\end{equation}
or, for the suggested moment of inertia ratio,
\begin{equation}
\mathcal{B} >  5\times 10^{-5} {m_\p^* \over m_\p} \, .
\end{equation}
This constraint is not very severe. In particular, the canonical value
$\mathcal{B} \sim 10^{-4}$ \citep{ALS84,1991ApJ...380..530M,ASC06}
lies within the required range. However, if we use the contraint of a
spin-up of the order of a day suggested by a partially resolved glitch
in the CRAB pulsar \citep{1992Natur.359..706L}, then the mutual
friction parameter would be constrained to being weaker than
$\mathcal{B}\sim 10^{-10}$.  This result would not accord well with
our current mutual friction models, likely illustrating our level of
ignorance about the relevant physics.

In principle, we would now want to consider the case when the
superfluid component represents only the free neutrons in the
crust. That is, when we have $I_\n /I \approx 10^{-2}$
\citep{1999PhRvL..83.3362L}. However, the model that we have developed
does not immediately apply to this situation. This is obvious since we
have assumed that two-fluid hydrodynamics applies throughout the
system. In order to adress the crust superfluid problem we would have
to add a component representing the single fluid core, and ensure that
it is coupled to the two-fluid region in a suitable way.  In
particular, this core component would affect
\eqref{eq:TotMomTwo}. This generalisation is complicated by the fact
that we would have to add appropriate boundary conditions at the
interfaces.  As our main aim is to compare the averaged model to the
detailed hydrodynamics we will leave consideration of models with
distinct superfluid regions for future work.

%%%%%%%%%%%%%%%%%%%%%%%%%%%%%%%%%%%%%%%%%%% SEC %%%%%%%%%%%%%%%%%%%%%%%%%%%%%%%%%%%%%
\subsection{Energetics}

In the next section we will consider the hydrodynamics associated with
a (core) glitch event.  A key motivation for this discussion is the
need to understand the actual details of how a macroscopic glitch is
triggered (presumably through a large-scale instability) and how the
system evolves once vortices become unpinned.  By modelling the
required hydrodynamics we hope to understand the nature of glitches
better. We should, for example, be able to establish to what extent
the simple ``bulk model'' we have discussed represents the behaviour
of a true two-fluid system. We can also address other interesting
questions, concerning for example the modes of oscillation that are
excited in a glitch. This is an interesting question to ask because,
first of all, there may be additional variability in the glitch event
and, secondly, the fluid oscillations may be associated with
gravitational-wave emission.  It is obviously relevant to try to
understand the nature of this gravitational-wave signal and estimate
its amplitude. One should probably not expect glitching pulsar to be
supreme gravitational-wave sources, but these estimates are
nevertheless interesting.  Most importantly, since glitches are common
in young pulsars (and magnetars) it may be ``reasonable'' to assume
that the corresponding level of energy is associated with regular
dramatic events in a neutron star's life.

Let us therefore consider the energetics of the problem. In past
studies, it has been common to estimate the energy available for
radiation based on a single component model. In that case the total
kinetic energy and angular momentum are (obviously) given by
\begin{equation}
E = { 1 \over 2} I \Omega^2 \ , \qquad \mbox{and } \qquad J = I\Omega
\, .
\end{equation}
Assume that a glitch of size $\Delta \Omega$ results from a change in
the moment of inertia $\Delta I$. This would represent a ``starquake''
in an elastic star. Then, assuming that the total angular momentum is
conserved, it is easy to show that the available energy is
\begin{equation}
\Delta E_1 \approx { 1 \over 2} I \Omega \Delta \Omega \, .
\end{equation}
As discussed by, for example, \citet{2001PhRvL..87x1101A} this
estimate suggests that pulsar glitches may be of interest for future
generations of gravitational-wave astronomers. However, if we consider
the two-component model we get a rather different picture. In this
case, for constant~$I_\x$, the conservation of angular momentum in the
glitch leads to
\begin{equation}
\Delta \Omega_\n = - { I_\p \over I_\n} \Delta \Omega_\p
\label{eq:domN} \, .
\end{equation}
That is, the superfluid (neutrons) spin down as the crust (protons)
spin up.  Estimating the available energy, we find that
\begin{equation}
\Delta E_2 \approx { 1 \over 2} I_\p (\Delta \Omega)^2 \, .
\end{equation}
Here, $\Delta \Omega = \Delta \Omega_\p$ and we have assumed that
$I_\p \ll I_\n$. For typical parameters, $I_\p / I \approx 0.1$ and
$\Delta \Omega/\Omega \approx 10^{-6}$, we see that
\begin{equation}
\Delta E_2 \sim 5 \times 10^{-8} \Delta E_1 \, .
\end{equation}
In other words, in the two-component model the energy available for
radiation is much smaller than in the starquake case. Even though it
is not clear how the estimate will change if we account for the energy
radiated as heat, changes in internal and potential energy etcetera,
it is clear that the result is rather pessimistic.  If the estimate is
taken seriously, and glitches really represent a transfer of angular
momentum as in the two-fluid model, then the gravitational-wave signal
from a pulsar glitch is unlikely to be detected by any future
generation of detectors. Of course, our level of understanding of this
problem is still rudimentary. We need to improve our models
considerably if we are to make more reliable estimates.  The
simulations that we will now discuss provide an important step in this
direction.

%%%%%%%%%%%%%%%%%%%%%%%%%%%%% SEC %%%%%%%%%%%%%%%%%%%%%%%%%%%%%%%%%%%%%%%%
\section{Hydrodynamics model}

The ``bulk'' model that we have discussed so far is able to describe
some key properties of glitches. However, the model has obvious
restrictions.  In particular, it does not provide any information
whatsoever about the actual hydrodynamics of the event. By focusing on
solid body motion we are obviously considering only the averaged
dynamics. In principle, one would expect the result to be relevant
when the dynamics is much slower than, say, the speed of sound in the
fluid. However, it is clear that we need to move beyond the averaged
model if we want to understand issues like the trigger mechanism for
glitches, consider potential gravitational-wave signals etcetera. It
is thus natural to consider the hydrodynamical aspects of the glitch
problem. To do this, we have extended the numerical code that was
recently developed by~\citet*{2009MNRAS.396..951P} [see
\citet{2005ApJ...635.1224P,2006ApJ...644L..53P,2009ApJ...701L..75P}
for a parallel effort].  Within the two-fluid framework, we evolve
perturbations of rotating, superfluid Newtonian stars in time. The new
version of the code includes the effects of mutual friction and the
perturbed gravitational potential (i.e. we are not working in the
Cowling approximation).  We initiate the time evolution with suitable
conditions that mimic a pre-glitch configuration, where the two fluids
rotate uniformly with different velocities.  Assuming that the vortex
pinning that is required to reach this state is instantaneously
broken, we can evolve the system. This allows us to determine the
mutual friction damping, extract the associated gravitational signal
and infer the oscillation modes that are excited during a
``glitch''. In particular, we can test the analytical formula for the
global spin-up timescale $\tau$, equation~(\ref{relax}).

The main motivation for our perturbative treatment is to consider
stellar models where the relative velocity lag between neutrons and protons is
very small as a
deviation from stationary equilibrium
configurations. These background models, which are such that the two
fluids co-rotate, are in $\beta$-equilibrium and coexist throughout
the star's volume, can be constructed by extending the standard
self-consistent field method of~\citet{Hachisu1986ApJS}. The details
of this method, and its application to superfluid stars can be found
in~\citet{2004MNRAS.347..575Y} and~\citet{2009MNRAS.396..951P}. In our
current models the crust is neglected (for simplicity). Our aim is to
continue to add key physics to the model, and we plan to consider
crustal effects in future work.

Non-corotating configurations can be determined using the perturbative
approach developed by~\citet{2004MNRAS.347..575Y}, where the
deviations from co-rotation are numerically computed by means of a
variation of the self-consistent field method. We extend this
numerical approach to study different superfluid equations of state
and generate rotating stellar sequences with constant mass. These
non-corotating deviations are then implemented in the hydrodynamical
code as initial conditions and evolved in time with a system of
perturbation equations. This system is formed by the linearised
versions of the two-fluid mass conservation equations, the two
Euler-type equations~(\ref{eq:EulMF}) and the Poisson equation for the
gravitational potential.  Technical details on the construction of the
initial data and the time domain numerical code will be provided
elsewhere~\citep{Passamonti-prep}. Here, we report only results that
can be directly compared with the analytic expressions from the
previous sections.

%------------------------------TAB. 1------------------------------------------%
\begin{table}
\begin{center}
\caption{\label{tab:back-models} This table provides the main
  parameters of the corotating background models for both the A and C
  sequences.  The first column labels each model, while the second and
  third columns give, respectively, the ratio of polar to equatorial
  axes and the angular velocity of the star. In the fourth column, the
  rotation rate is compared to the Kepler velocity $\Omega_K$ that
  represents the mass shedding limit. The ratio between the rotational
  kinetic energy and gravitational potential energy $T/|W|$ is given
  in the fifth column. In the sixth and seventh columns we show the
  moment of inertia of the proton and neutron fluids, respectively,
  while in the eighth column we provide the stellar mass. All
  quantities are given in dimensionless units, where $G$ is the
  gravitational constant, $\rho_0$ represents the central mass density
  and $R_{eq}$ is the equatorial radius.}
\begin{tabular}{ c c  c c c c c c  }
\hline
 Model & $ R_p / R_{eq} $  &  $ \Omega / \sqrt{G\rho_0}$ & $\Omega / \Omega_K$  & $ T/|W| \times 10^{2}$
  & $ I_\p / (\rho_0 R_{eq}^5)$ & $ I_\n / (\rho_0 R_{eq}^5)$ & $ M / (\rho_0 R_{eq}^3)$   \\
\hline
 A0 &  1.00000   &   0.00000   &  0.00000  &    0.00000  &  0.03328 & 0.29951  &  1.2732   \\
 A1 &  0.99792   &   0.05913   &  0.08156  &    0.05802  &  0.03319 & 0.29868  &  1.2701   \\
 A2 &  0.98333   &   0.16675   &  0.22999  &    0.38482  &  0.03253 & 0.29278  &  1.2479   \\
 A3 &  0.95000   &   0.28799   &  0.39627  &    1.16918  &  0.03102 & 0.27915  &  1.1967   \\
 A4 &  0.93333   &   0.33081   &  0.45629  &    1.56885  &  0.03025 & 0.27224  &  1.1709   \\
 A5 &  0.90000   &   0.40268   &  0.55543  &    2.38295  &  0.02869 & 0.25822  &  1.1186   \\
\\
 C0 &  1.00000   &   0.00000   &  0.00000  &    0.00000  &  0.01906 & 0.24657  &  1.0826  \\
 C1 &  0.99792   &   0.05586   &  0.08403  &    0.04561  &  0.01900 & 0.24584  &  1.0798  \\
 C2 &  0.98333   &   0.15764   &  0.23716  &    0.36682  &  0.01858 & 0.24064  &  1.0601  \\
 C3 &  0.95000   &   0.27145   &  0.40837  &    1.11303  &  0.01760 & 0.22861  &  1.0146  \\
 C4 &  0.93333   &   0.31246   &  0.47006  &    1.49236  &  0.01711 & 0.22251  &  0.9915   \\
 C5 &  0.90000   &   0.38006   &  0.57177  &    2.26285  &  0.01610 & 0.21009  &  0.9447  \\
\hline
\end{tabular}
\end{center}
\end{table}
%------------------------------------------------------------------------------%

In a two-fluid model, the relative motion between protons and neutrons
can be approximately decomposed in co- and counter-moving
components. With an appropriate choice of perturbation variables, we
can study the effects of these two degrees of freedom on the
oscillation spectrum and the factors that generate their
coupling~\citep{2009PhRvD..79j3009A, 2009MNRAS.396..951P}.

The perturbation equations must be completed with an equation of
state (EoS), which can be described by the energy
functional~(\ref{eq:intE}):
\begin{equation}
E = E \left(\rho_\n, \rho_\p , w_{\n \p}^2 \right) \label{eq:EoSfunc} \,
,
\end{equation}
where we have replaced the number densities $n_\x$ with the mass
densities $\rho_\x$ (assuming for simplicity that the neutron and
proton masses are equal, $m_\p = m_\n$). When the relative velocity
between the two fluids is small, equation~(\ref{eq:EoSfunc}) can be
expanded in a Taylor series~\citep{2002A&A...381..178P,
2009MNRAS.396..951P}.  Then we have
\begin{equation}
E = E_{0} \left(\rho_\n, \rho_\p \right) + \alpha_0 \left( \rho_\n,
\rho_\p \right) w_{\n \p}^2 + \mathcal{O}\left(w_{\n \p}^4\right)  \label{eq:EoSbulk} \, ,
\end{equation}
where the entrainment function, $\alpha_0$, and the bulk equation of
state, $E_0$, can be independently specified on the corotating
background, where $w^{i}_{\p\n} = 0$. In this paper, we consider two
sets of polytropic superfluid equations of state already used
by~\citet{2009MNRAS.396..951P}. Despite their simplicity, these models
are very useful for investigating the effects of the different
physical parameters on the oscillation dynamics. In particular, one of
these two classes of equation of state describes models with
composition gradients that couple the co- and counter-moving degrees
of freedom.

The first equation of state is given by the following expression:
\begin{equation}
E_{0} = \frac{K}{1-\left( 1 + \sigma \right) x_\p} \rho_\n ^2 -
\frac{2 K \sigma }{1-\left( 1 + \sigma \right) x_\p} \rho_\n \rho_\p +
\frac{ K \left[ 1 + \sigma - \left( 1 + 2 \sigma \right) x_\p \right]
} {x_\p \left[ 1-\left( 1 + \sigma \right) x_\p \right] } \rho_\p^2
\label{eq:EoS1} \, ,
\end{equation}
where $K$ and the proton fraction $x_\p$ are taken to be constant.  As
discussed by \citet{2002A&A...381..178P}, the parameter $\sigma$ is
related to the ``symmetry energy'' of the EoS.  For this model, we
have constructed a sequence of co-rotating axisymmetric configurations
which do \underline{not} have composition gradients. These
non-stratified stars correspond to the A models used
by~\citet{2009MNRAS.396..951P}. Table~\ref{tab:back-models} gives the
main quantities of some rotating models that we consider in this
paper. Details for more rapidly rotating models can be found
in~\citet{2009MNRAS.396..951P}.

The second analytical equation of state, that describes stratified
superfluid stars, can be written~\citep{2002A&A...393..949P,
2002PhRvD..66j4002A, 2009MNRAS.396..951P}:
\begin{equation}
E_{0} = k_{\n} \, \rho_\n^{\gamma_n} + k_{\p} \, \rho_\p^{\gamma_\p}  \label{eq:EosPR} \, .
\end{equation}
In this paper we determine a sequence of rotating stars, whose
non-rotating member is Model III of~\cite{2002A&A...393..949P}. For
our polytropic models we use $\gamma_\n = 1.9$ and
$\gamma_\p = 1.7$, while the coefficients~$k_\x $ are given, in units
$G R_{eq} \rho_{0}^{2-\gamma_x}$, by $k_\n=0.682$ and $k_\p=3.419$,
respectively. Recall that $G, R_{eq}$ and $\rho_0$ are the
gravitational constant, the equatorial radius and the central mass
density, respectively. Imposing $\beta$-equilibrium on the corotating
background model, we can determine the proton fraction as
~\citep{2002A&A...393..949P, 2009MNRAS.396..951P}:
\begin{equation}
x_\p = \left[ 1 + \frac{ \left( \gamma_\p k_\p \right) ^{N_\p}
}{\left( \gamma_\n k_\n \right) ^{N_\n}} \, \tilde \mu ^{N_\n-N_\p} 
\right]^{-1}   \, ,\label{eq:xp}
\end{equation}
where $N_\n$ and $N_\p$ are the neutron and proton polytropic indices of
the EoS, defined as $N_\x = \left( \gamma_\x - 1 \right)^{-1}$.
For this model, a sequence of rotating stars can then be determined
via the self-consistent field method~\citep{2009MNRAS.396..951P,
Passamonti-prep}. We refer to these stars as models C, in order to
distinguish them from the models B used
by~\cite{2009MNRAS.396..951P}. See Table~\ref{tab:back-models} for
some of the rotating configurations for this model.

%%%%%%%%%%%%%%%%%%%%%%
\subsection{Glitch initial data} \label{sec:ID}

According to the two-fluid model, a glitch brings a star with an
initial velocity lag between protons and neutrons to a co-rotating
configuration. The observed glitch jump is very small, $\Delta \Omega
/ \Omega \ll 1$, and can be associated with angular momentum transfer
from the superfluid neutrons to the proton component.  Regardless of
the physical mechanism that generates the original velocity
difference, we can describe the initial conditions for a glitch as an
axisymmetric configuration where protons and neutrons rotate with a
small relative velocity.  If we assume that this relative rotation is
aligned with the ``background'' rotation axis, we can use the
perturbative approach developed by~\cite{2004MNRAS.347..575Y} to
determine the difference between the initial non-corotating
configuration and the final corotating background.

As we have already discussed, the observed variation of the star's
 angular velocity during a glitch can be associated with the proton
 velocity. In fact, due to the magnetic field coupling between the
 crust and core protons, we can assume that all charged particles
 corotate. Meanwhile, the velocity of the superfluid neutrons must be
 determined from the conservation laws.  If we assume that the angular
 momentum of the two-fluid system is conserved during a glitch, then
 the initial relative amplitude between the proton and neutron
 perturbations can be determined from
\begin{equation}
\sum_{\x} \Delta J_\x = 0  \label{eq:dJ} \, ,
\end{equation}
where the perturbed angular momentum of the x component is given by
\begin{equation}
\Delta J_\x = I_\x \Delta \Omega_\x + \Delta I_\x \, \Omega \, ,
\end{equation}
and $I_\x$ is the moment of inertia of the x fluid in the corotating
configuration, which rotates with angular velocity $\Omega$.  We
determine the perturbation $\Delta I _\x$ from
\begin{equation}
\Delta I_\x = \int_{0}^{\mtb{r}} \delta \rho_\x \, (r' \sin \theta )^2 d\mtb{r'} \, ,
\end{equation}
where $\delta \rho_\x$ is the Eulerian perturbation of the mass
density. From equation~(\ref{eq:dJ}) we then have
\begin{equation}
I_\n \Delta \Omega_{\n} +  I_\p  \Delta \Omega_{\p}
+  \left( \Delta I_\n + \Delta I_\p \right) \Omega = 0  \label{eq:arel} \, ,
\end{equation}
which leads to the following expression for the initial fluid angular
velocities:
\begin{eqnarray}
\frac{\Delta \Omega_\p }{\Omega}  & = &  \left. \frac{ \Delta \Omega }{ \, \, \Omega }\right|_{obs} \, , \label{eq:arel1}\\
\Delta \Omega_{\n} & = & - \frac{1 }{I_\n }  \left(  I_\p\Delta \Omega_{\p}
+ \Delta I \, \Omega \right)   \label{eq:arel2} \, ,
\end{eqnarray}
where $\Delta I = \Delta I_\p + \Delta I_\n$.

For any corotating background we can determine the non-corotating
corrections to the mass density $\delta \rho_\x$, the chemical
potential $\delta \tilde \mu_\x$ and the gravitational potential
$\delta \phi$ via the~\citet{2004MNRAS.347..575Y} approach,
see~\cite{Passamonti-prep} for more details. In a spherical coordinate
basis, the initial velocity fields are then given by
\begin{equation}
\qquad \delta v_\x^{i} = - \Delta \Omega_\x \, \varphi^{i}   \label{eq:vphi} \, .
\end{equation}
This initial data is close to that considered in the bulk dynamics
model of the previous section. Hence, we can meaningfully compare the
results of our time evolutions to the averaged results.  This
comparison will give us a better idea of the validity of the
solid-body approach. Of course, by solving the hydrodynamics problem
we will be able to proceed beyond the averaged model and discuss other
interesting issues.

%-----------------FIG-----------------------------
\begin{figure}
  \centering
    \includegraphics[height = 70mm]{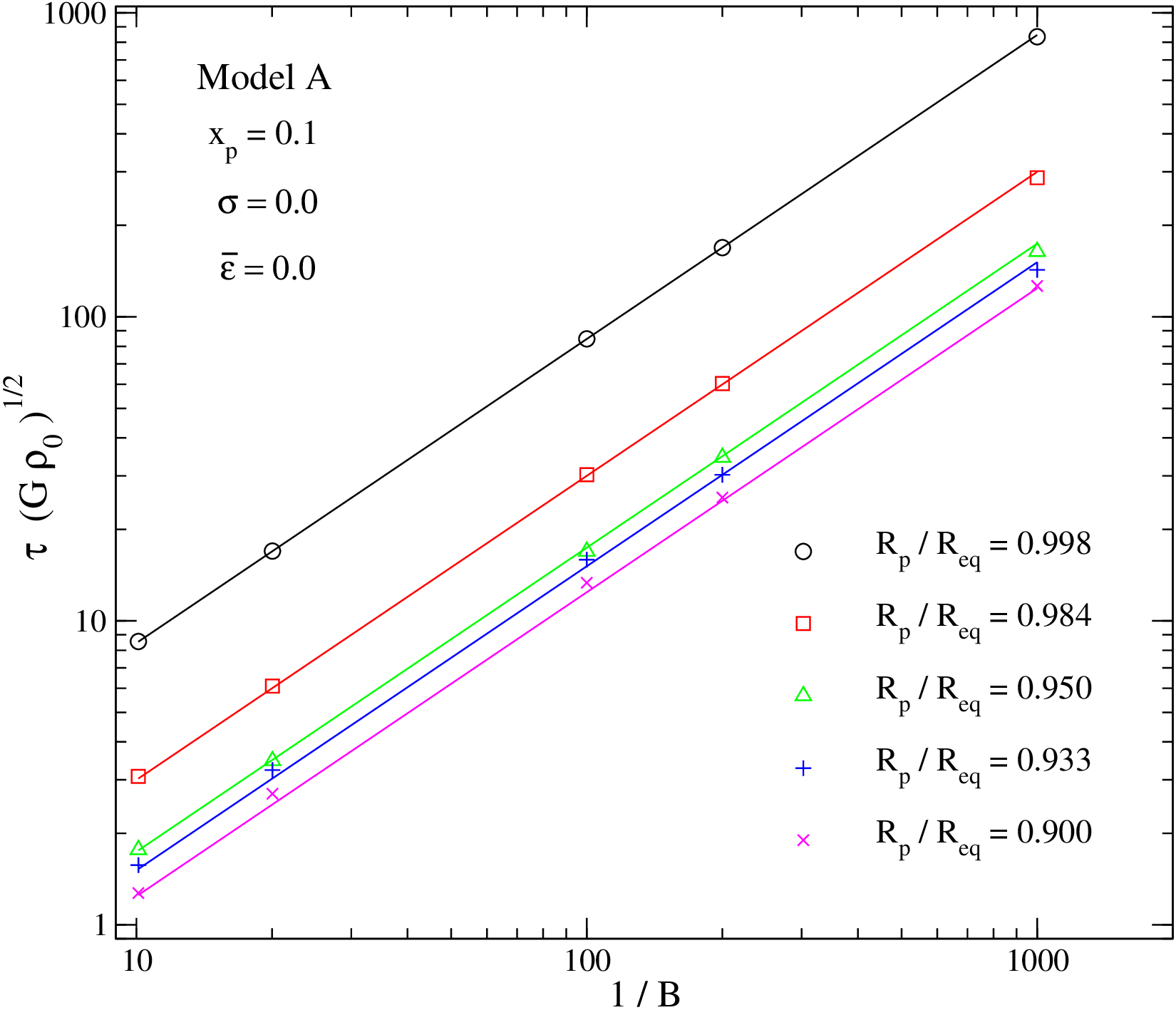}%{fig-tau-B.eps}
    \includegraphics[height = 69.6mm]{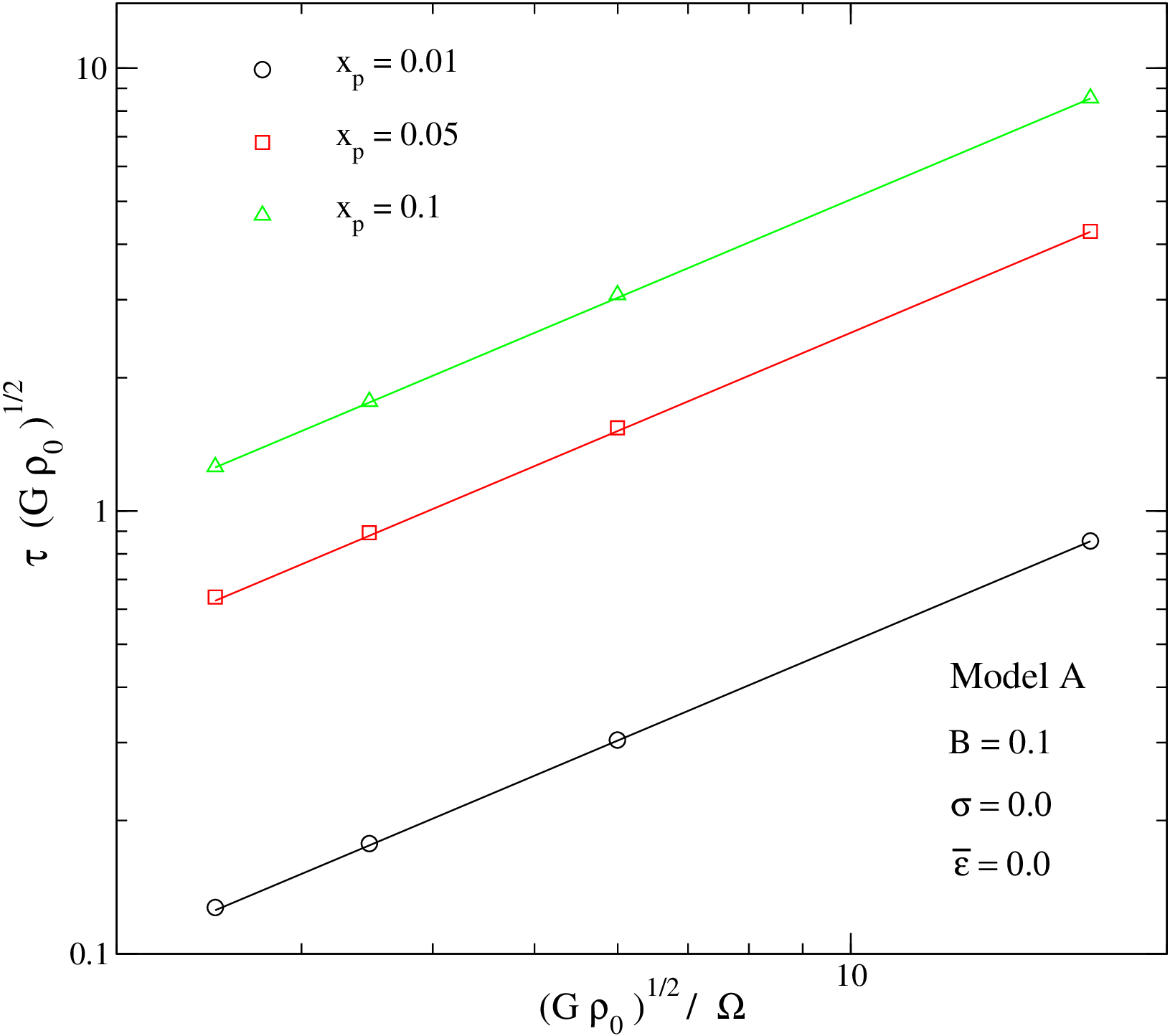}%{fig-tau-omega.eps}
  \caption{This figure shows the spin-up time $\tau$ as a function of
  the inverse of the mutual friction parameter $\mathcal{B}$ (left
  panel) and the background rotation rate of the star $\Omega$ (right
  panel) for the sequence of rotating models A.  The solid lines show
  the behaviour predicted by equation~(\ref{relax}), while the symbols
  (see legend) represent the values of the spin-up time extracted from
  the hydrodynamical simulations.  The physical quantities are given
  in dimensionless units by using the gravitational constant $G$ and
  the central mass density $\rho_0$. The axes use logarithmic
  scales. All the five models A1-A5 shown in this figure have both
  vanishing symmetry energy term $\sigma$ and entrainment parameter
  $\bar \veps$. In the left panel, the proton fraction is fixed to
  $x_\p=0.1$ and the mutual friction is varied. Meanwhile, in the
  right panel we show three sequences of rotating stars with the same
  mutual friction strength $\mathcal{B} = 0.1$, but with three
  different values of proton fraction, namely $x_\p = 0.1, 0.05$ and
  $x_\p=0.01$. In all cases, the numerical values of spin-up time show
  a good agreement with the analytical result.}
  \label{fig:tau1}
\end{figure}
%----------------------------------------------

%%%%%%%%%%%%%%%%%%%%%%%%%%% SEC %%%%%%%%%%%%%%%%%%%%%%%
\subsection{Spin-up time}

Let us first compare the mutual friction damping extracted from the
time-evolution to the body averaged analytical
formula~(\ref{relax}). Considering the sequence of non-stratified
models A, we test the dependence of $\tau$ on the mutual friction
strength $\mathcal{B}$, the background angular velocity $\Omega$, the
moment of inertia ratio $I_\p/I$ and the entrainment parameter~$\bar
\veps$. For the corotating background A models given in
Table~\ref{tab:back-models}, we determine the initial condition for
the time evolution code that corresponds to a ``typical'' glitch jump
$\Delta \Omega / \Omega = 10^{-6}$. The related neutron velocity lag
is then given by equation~(\ref{eq:arel2}). It is worth noting that
the method used to construct the initial data is linear in the two
fluid velocities. This means that the results can be rescaled to other
values of the glitch size.

If we exclude the mutual friction term, the numerical evolutions
preserve the initial lag between protons and neutrons. At the same
time, the initial conditions excite low level oscillations. This is
expected as we are mapping a non-corotating axisymmetric configuration
onto an axisymmetric corotating background.  When mutual friction is
switched on, the counter-moving motion is damped, leading to a
corotating configuration on a timescale of order $\tau$. We determine
the spin-up timescale by monitoring the $\varphi$ component of the
velocity difference $w_{\p\n}^i$, assuming that it depends on time as
$w_{\p\n}^{\varphi} = w_{\p\n}^{\varphi}\left(t=0\right) \,
e^{-t/\tau}$. By taking the natural logarithm, we can extract $\tau$
from a linear fit of the time evolved data.

\begin{figure}
  \centering
    \includegraphics[height = 70mm]{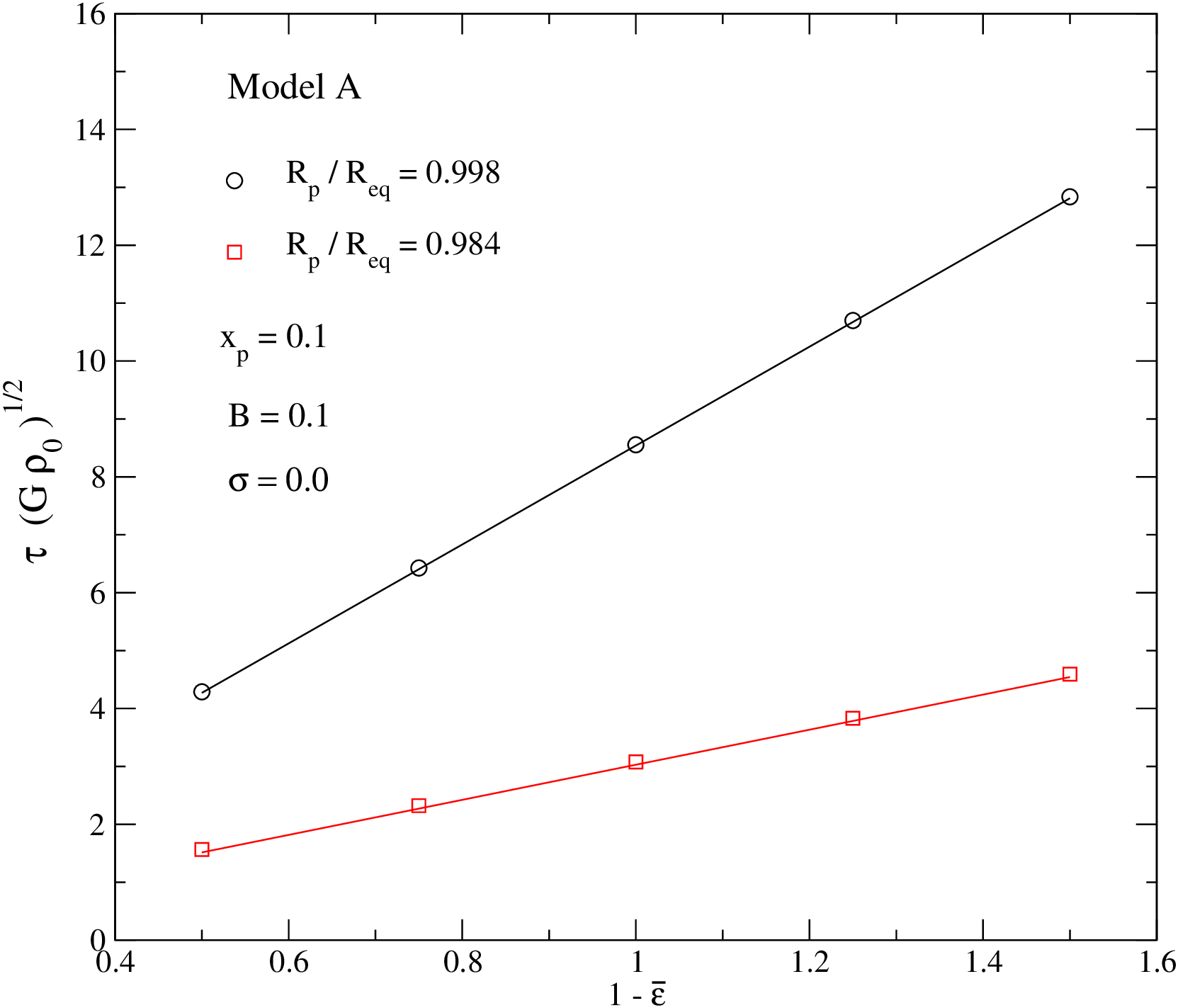}%{fig-tau-eps.eps}
    \includegraphics[height = 70mm]{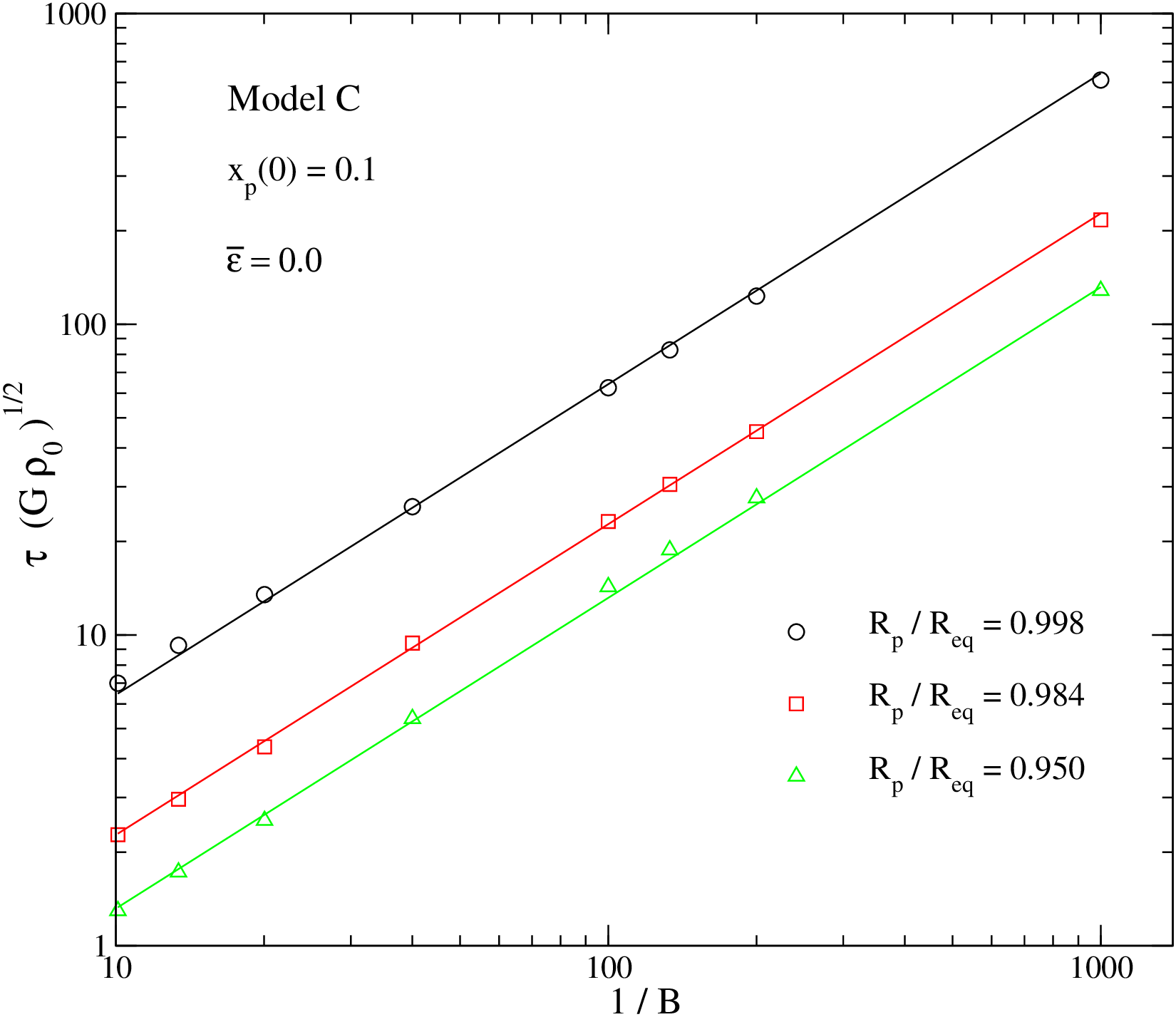}%{fig-tau-C-model-B.eps}
  \caption{In this figure we compare the numerical spin-up time
  $\tau$ with the analytical formula~(\ref{relax}). The values
  extracted from the numerical code are shown as open symbols (see
  legend), while the solid lines denote the analytical $\tau$. In the
  left panel, we consider the two models A1 and A2, with proton
  fraction $x_\p =0.1$, vanishing symmetry energy $\sigma = 0$ and
  constant mutual friction parameter $\mathcal{B} = 0.1$. The
  dependence of the numerically determined $\tau$ on the entrainment
  parameter agrees very well with the analytical result. In the right
  panel, we show (on a log-log scale) the damping time $\tau$ as a
  function of $\mathcal{B}^{-1}$ for the three models C1-C3 with
  vanishing entrainment. The agreement between the numerical and
  analytical spin-up times is still good, although less accurate
  than for the A models. See Fig.~\ref{fig:tau3} and the main text for
  further discussion.}
  \label{fig:tau2}
\end{figure}

In Figs.~\ref{fig:tau1} and~\ref{fig:tau2}, we show the agreement of
the numerical results with the analytical formula~(\ref{relax}). In
the left panel of Fig.~\ref{fig:tau1}, we consider several rotating A
models with vanishing symmetry energy and entrainment and with
constant proton fraction $x_\p = I_\p / I = 0.1$. The models have
different mutual friction strength, controlled by the parameter
$\mathcal{B}$ that we take to be constant throughout the star. The
expected linear dependence of $\tau$ on the inverse of the mutual
friction parameter is confirmed by the hydrodynamical simulations. In
the right panel of Fig.~\ref{fig:tau1}, we fix instead the mutual
friction strength, $\mathcal{B}=0.1$, a rather large value, and study
three sequences of rotating models with different proton fraction,
respectively $x_\p = 0.01, 0.05$ and 0.1. The mutual friction damping
time exhibits the expected linear dependence on $\Omega^{-1}$.  For
models A1 and A2 we test also the dependence of $\tau$ on the
entrainment parameter~$\bar \veps$, see Fig~\ref{fig:tau2}. In this
case, the other stellar parameters are, respectively, $x_\p = 0.1$,
$\mathcal{B}=0.1$ and $\sigma=0$. The linear dependence on $1-\bar
\veps$ is clearly confirmed by the evolutions. According to
equation~(\ref{relax}), the damping time should not depend
(explicitly) on the symmetry energy, $\sigma$. We have carried out
simulations with different $\sigma$ to confirm this result.

\begin{figure}
  \centering
    \includegraphics[height = 70mm]{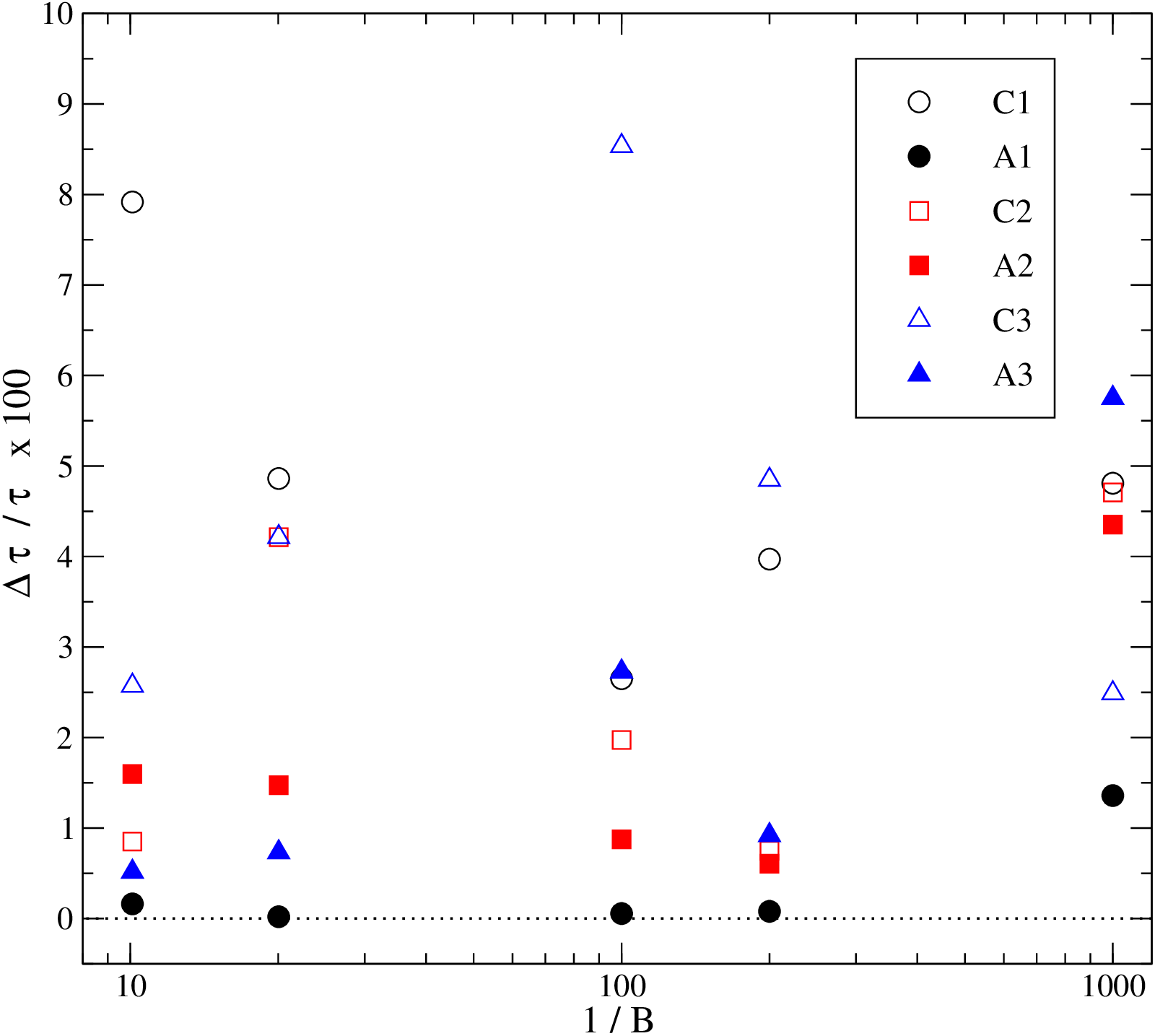}%{fig-perc.eps}
  \caption{In this figure, we estimate the agreement between the
  analytical spin-up time $\tau$ from equation~(\ref{relax}) and
  the values extracted from the hydrodynamical code.  We show the
  relative deviation~$\Delta \tau / \tau$ as a function of the inverse
  of the mutual friction parameter $\mathcal{B}$. The horizontal axis
  is on logarithmic scale. The results for models A and C are shown as
  filled and open symbols, respectively~(see legend for details).
    \label{fig:tau3}}
\end{figure}

Next, we consider the stratified models C. The aim is to establish to
what extent equation~(\ref{relax}) still provides accurate results for
the spin-up time. On the one hand, one may not expect this to be
the case since the various parameters in the model are no longer
uniform. On the other hand, the simple prescription could still work
provided that the parameters are interpreted in a body-averaged
sense. We consider initial configurations with $\Delta \Omega / \Omega
= 10^{-6}$ and determine the non-corotating corrections using the
method discussed in Section~\ref{sec:ID}. For the three rotating
models C1-C3, the dependence of the numerical damping time on the
mutual friction parameter $\mathcal{B}$ is shown in the right panel of
Fig.~\ref{fig:tau2}.  Again, the results agree well with the
analytical formula~(\ref{relax}). However, a closer examination of the
data reveals that equation~(\ref{relax}) is more accurate for the
non-stratified A models.  For the first three rotating models of the
two sequences A and C, we show in Fig.~\ref{fig:tau3} the relative
deviation between the numerical and analytical spin-up time for
different mutual friction strengths.  Comparing models with the same
axis ratio, the error is generally smaller for models A (filled
symbols).

In general we find that the agreement between the numerical and
analytical spin-up time is better for slowly rotating models that
have strong mutual friction. In these cases, the damping of
$w_{\p\n}^i$ is less contaminated by the excitation of axisymmetric
oscillations, and the numerical extraction of $\tau$ is more accurate.

%-----------------FIG-----------------------------
%----------------------------------------------

%%%%%%%%%%%%%%%%%%%%%%%%%%% SEC %%%%%%%%%%%%%%%%%%%%%%%
\subsection{Gravitational Waves}

So far we have discussed ``global'' dynamics, e.g. how the two
components in the system relax to a co-rotating configuration due to
the mutual friction.  We now turn to the actual hydrodynamics, and
consider to what extent this kind of glitch event radiates
gravitational waves. We have already established that these events are
unlikely to be strong emitters of gravitational radiation. However, it
seems inevitable that they should radiate at some level and it is
important to establish what that level may be. In particular, since
there are a number of glitching pulsars in the Galaxy. The energy
involved in glitches indicates how energetic ``typical'' events in a
mature neutron star may be. In that way, these events provide
interesting benchmarks for gravitational-wave modelling. It is,
however, important to state already from the outset that we are not
expecting the mechanism that we are considering here to lead to
detectable signals.  This is essentially because of the high level of
symmetry in the initial and final configurations, and the fact that we
are assuming global vortex unpinning. The result may be quite
different if we were to model localized unpinning events. Although of
great interest, this problem is unfortunately beyond the reach of our
current computational technology.

We will focus on the gravitational signal associated with the $l=2$
axisymmetric oscillations that are excited in our glitch evolutions.
At the linear perturbation level, the initial data excites a number of
the neutron star's oscillation modes. Hence, a key question concerns
which modes we expect to be present in the gravitational signal. For a
single fluid star, the general mode classification is based on the
main restoring force that acts on the displaced fluid
elements~\citep{Cowling:1941co}. In this work we consider ``acoustic
modes'' that are mainly restored by pressure variations.  Since any
perturbation of a spherical star can be decomposed in vector
harmonics, an oscillation mode can be labeled by the harmonic indices
$(l,m)$ associated with the spherical harmonics
$Y_l^m(\theta,\phi)$. This description can be extended to rotating
stellar models, as the modes can be tracked back to the non-rotating
limit.  For any value of $(l,m)$, the oscillation modes can be ordered
by the number of radial nodes in their eigenfunctions.  For acoustic
modes, we then have the fundamental mode ${}^l \textrm{f}$ with no
nodes and the series of pressure modes ${}^l\textrm{p}_i$ with $i$
nodes.

Our glitch model leads to axisymmetric initial data. Therefore, we can
only excite the family of axisymmetric modes, with $m=0$.  The
quadrupole, $l=2$, oscillations are expected to be dominant in the
gravitational signal. However, in rotating stars the
gravitational-wave spectrum can also contain $l=0$ ``quasi-radial''
oscillation modes.  In the non-rotating limit, these become purely
radial modes, which do not generate gravitational radiation. The
quasi-radial fundamental mode will be denoted by F and its $i$
overtones by $\textrm{H}_i$.

In a two-fluid model the oscillation spectrum is richer, as the
displaced fluid elements can now oscillate in phase and counter-phase.
The comoving degrees of freedom generate oscillation modes that are
similar to those of a single fluid star, and are referred to as
``ordinary modes''. The counter-moving degree of freedom produce a new
class of modes known as ``superfluid modes''.  In our discussion,
ordinary and superfluid modes will be labeled with an upper index, for
instance the $l=2$ fundamental ordinary mode will be expressed as
${}^2 \textrm{f}^{\hspace{0.5 mm} \rm{o}}$, while the superfluid mode
as ${}^2 \textrm{f}^{\hspace{0.6 mm} \rm{s}}$.

 In a two-fluid star, the gravitational radiation is entirely
 generated by the co-moving degree of
 freedom~\citep{2009PhRvD..79j3009A}. We determine the gravitational
 strain from the standard quadrupole
 formula~\citep{1980RvMP...52..299T}:
\begin{equation}
h_{+}^{20} = \frac{G}{c^4} \frac{1} {r} \frac{d^2 \mathcal{I}}{dt^2}^{20} \,  T_{\theta \theta} ^{E2,20}    \label{eq:hp} \, ,
\end{equation}
where the $(l,m)=(2,0)$ pure spin tensor harmonic is given by
\begin{equation}
 T_{\theta \theta} ^{E2,20} = \frac{1}{8} \sqrt{ \frac{ 15}{\pi }} \sin^2 \theta \, .
\end{equation}
The mass quadrupole moment for the two-fluid star is defined
by~\citep{2009PhRvD..79j3009A}:
\begin{equation}
\mathcal{I}^{20} = 8 \sqrt{ \frac{ \pi}{15} } \int d\mtb{r} \, \delta
\rho \, r^2 P^{20} = 8 \sqrt{ \frac{ \pi}{15} } \int d\mtb{r} \,
\left( \delta \rho_\n + \delta \rho_\p \right) r^2 P^{20} \, ,
\end{equation}
where $P^{20}$ is the associated Legendre polynomial:
\begin{equation}
P^{20} = \frac{ 3 \cos ^2 \theta - 1 }{2} \, .
\end{equation}
In equation~(\ref{eq:hp}), the numerical calculation of the second
order time derivative of the quadrupole moment may lead to
inaccuracies.  However, the gravitational-wave extraction can be
improved by using the dynamical equations and replacing the time
derivative by spatial derivatives~\citep{1990ApJ...351..588F}.  In our
perturbative analysis, we have found accurate results already with the
momentum-formula, where only first time derivatives appear. More
details and tests will be given by~\cite{Passamonti-prep}.

We rewrite equation~(\ref{eq:hp}) as follows:
\begin{equation}
h_{+}^{20} = \frac{G}{c^4} \frac{\sin^2 \theta} {r} \sum_{\x}
A_\x^{20}   \, ,\label{eq:hp3}
\end{equation}
where the quantity $A_\x^{20}$ is given by~\citep{Passamonti-prep}
\begin{equation}
A^{20}_\x \equiv 8 \pi \frac{d }{dt} \int_{0}^{\pi/2} \sin \theta d
\theta \int _{0}^{R} r^3 d r \rho_\x \left( \delta v_\x^{r} \, P^{20} +
r \frac{\delta v_\x^{\theta} }{2}   \frac{ \partial P}{\partial
\theta}^{\hspace{-0.05cm}20} \right) \, .
\label{eq:A20-mom}
\end{equation}
The energy radiated in gravitational waves can be determined from the
following equation~\citep{1980RvMP...52..299T}:
\begin{equation}
 E_{rad} = \frac{1}{32\pi} \frac{G}{c^5} \int_{-\infty}^{\infty}
\left| \frac{d^3 \mathcal{I}}{dt^3}^{\hspace{-0.04cm}20} \right| ^{2}
dt = \frac{2}{15} \frac{G}{c^5} \int_{-\infty}^{\infty} \left| \frac{d
A}{dt}^{\hspace{-0.05cm}20} \right| ^2 dt = \frac{16}{15}\pi^2
\frac{G}{c^5} \int_{0}^{\infty} \nu^2 \left| \hat{A}^{20} \right| ^2 d
\nu \, ,
\end{equation}
where $A^{20} = A^{20}_\n + A^{20}_\p $, and $\hat{A}^{20}$ is its
Fourier transform.  The energy spectrum is then given by
\begin{equation}
 \frac{ d E}{d \nu} = \frac{G}{c^5} \frac{16 }{15}\pi^2 \nu^2 \left|
 \hat{A}^{20} \right| ^2 \, ,
\end{equation}
and the characteristic gravitational-wave strain is defined
as~\citep{1998PhRvD..57.4535F}:
\begin{equation}
 h_c \left(\nu \right) \equiv \sqrt{\frac{G}{c^3}}  \frac{\sqrt{2}}{\pi} \frac{1}{d} \sqrt{ \frac{  d E}{d \nu} }
= \frac{G}{c^4} \sqrt{\frac{32}{15} } \, \frac{\nu}{d}  \, \left| \hat{A}^{20} \right|   \label{eq:hcdef} \, .
\end{equation}
In our analysis, we focus on the two rapidly rotating models A2 and
C2, which rotate at a significant fraction of the mass shedding
limit~$\Omega/\Omega_{K} = 0.23$~(see Table~\ref{tab:back-models} for
more details). For a typical neutron star with mass $M=1.4M_{\odot}$
and radius $R_{eq}=10~\textrm{km}$, the rotation period of the A2 and
C2 models is about $P\simeq 3~\textrm{ms}$.  These models are
therefore rotating much faster than all known glitching pulsars.  We
will demonstrate that the generation of gravitational waves is very
small even for these relatively rapidly rotating models. The result
can be considered as an upper limit for real glitching neutron
stars. Moreover, we will show that the gravitational signal of slower
rotating models can be determined by a simple rescaling of the A2 and
C2 signals.

%-----------------FIG-----------------------------
%----------------------------------------------

At the linear perturbation level, the entrainment parameter $\bar
\veps$ can be chosen independently from the background model.  Recent
work suggests that this parameter can assume values in the range $0.2
\leq \bar{\veps} \leq 0.8$~\citep{2008MNRAS.388..737C}.  The effect of
the entrainment on the oscillation spectrum has been extensively
studied by~\cite{2002A&A...393..949P} and
\cite{2009MNRAS.396..951P}. The oscillation modes that are mainly
affected by $\bar \veps$ are the so-called ``superfluid'' modes, which
are associated with the counter-moving degree of freedom. Simple
relations between the frequencies of the superfluid acoustic and
inertial modes and the entrainment parameter $\bar \veps$ have been
obtained by~\cite{2009PhRvD..79j3009A,
2009MNRAS.396..951P,2009MNRAS.tmp..951H}.  In this work we show
results only for the $\bar \veps = 0.5$ case, as apart from the
spectral properties discussed above, we did not find any qualitative
difference in simulations using other values of this parameter.

With the method discussed in Section~\ref{sec:ID}, we set up initial
data that mimic the configuration of an axisymmetric glitch. The
initial value for the proton and neutron angular velocity can be
determined from equations~(\ref{eq:arel1})-(\ref{eq:arel2}), once we
fix the glitch size $\Delta \Omega / \Omega$ and solve the stationary
equations for the background model. We will consider the case of a
large glitch, where $\Delta \Omega_\p / \Omega = 10^{-6}$.  This means
that, due to angular momentum conservation, the neutron fluid slows
down with $\Delta \Omega_\n / \Omega = - 1.11\times 10^{-7}$ for model
A2 and $\Delta \Omega_\n / \Omega = - 7.74\times 10^{-8}$ for model
C2. Note that we use a first order perturbative framework, where for a
given corotating background model the results of the time evolutions
are linear with respect to the parameter $\Delta \Omega_\p /
\Omega$. Hence, the gravitational-wave strain can be rescaled to any
desired glitch magnitude.  Furthermore, for slower rotating models
that have the same glitch size, $\Delta \Omega_\p / \Omega$, we expect
the perturbations and the gravitational strain to exhibit a quadratic
dependence on the background rotation rate $\Omega$. Our numerical
simulations reproduce this behaviour when the stars have small
rotational deformations, as in the case of the A1-A2 and C1-C2
models. Already for models A3 and C3, this scaling with $\Omega ^2$ is
less clear and obviously it is not expected to hold for more rapidly
rotating stars. In conclusion, from the evolutions of the A2 and C2
models, we can easily estimate the gravitational strain emitted by
other slowly rotating models with different glitch size and background
rotation.

\begin{figure}
  \centering
    \includegraphics[height = 71.2mm]{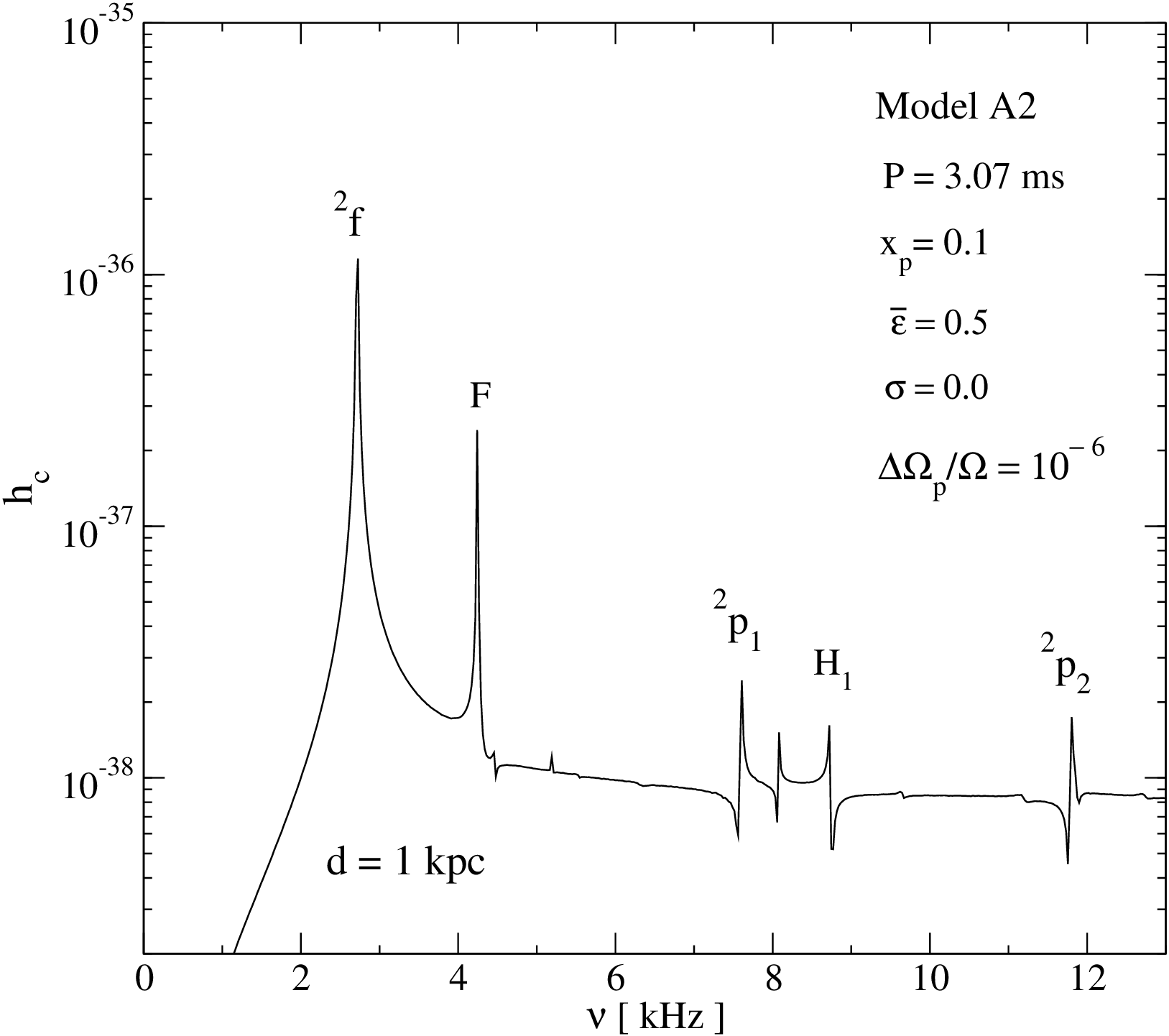}
    \includegraphics[height = 70mm]{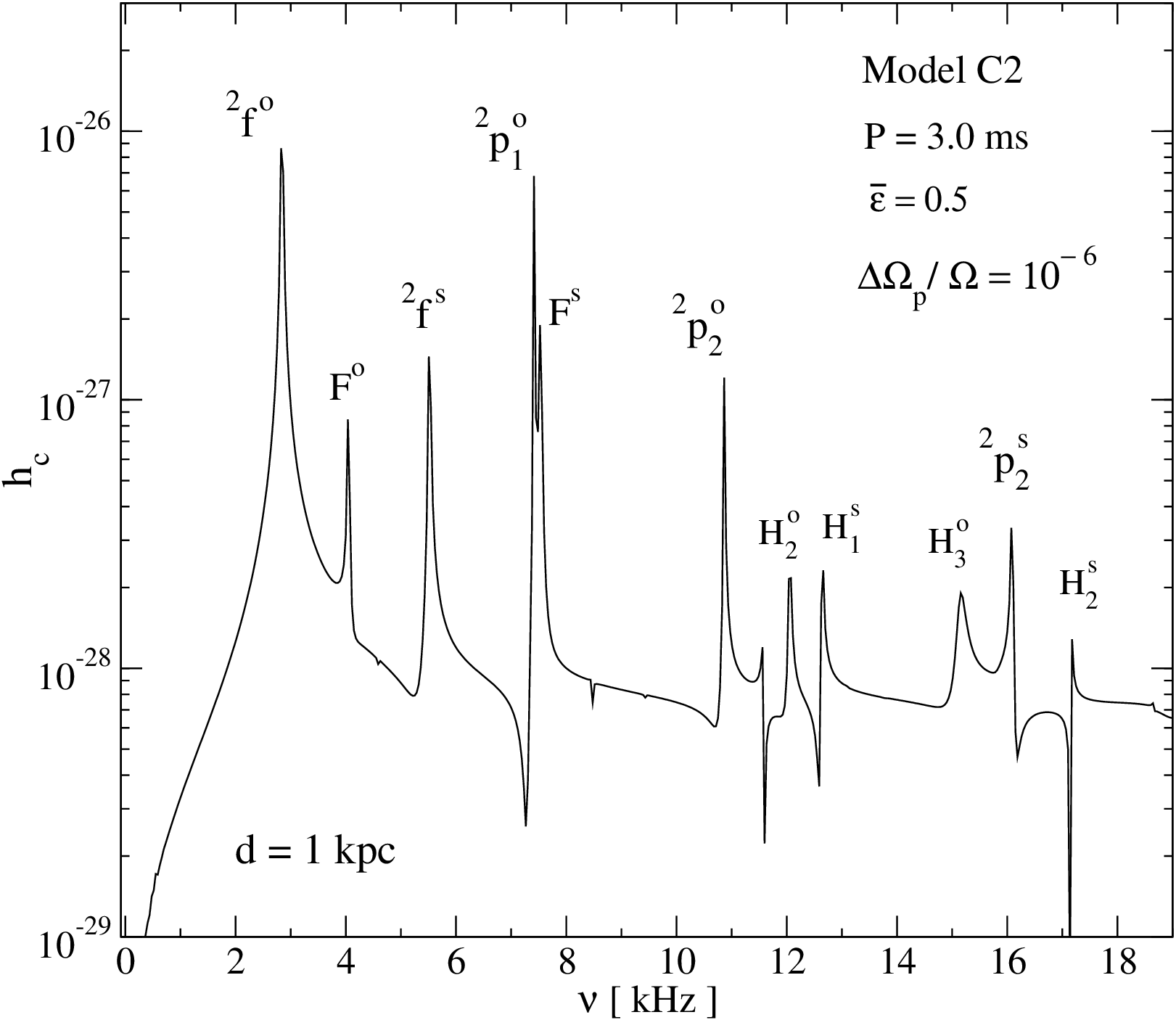}   
  \caption{This figure displays the gravitational-wave signal
  generated by our hydrodynamical glitch simulations for the two
  models A2 (left panel) and C2 (right panel).  On the horizontal and
  vertical axes we plot the oscillation frequencies and the
  characteristic strain extracted at a distance of $1~\rm{kpc}$ from
  the source.  We consider a neutron star with typical mass $M=1.4
  M_{\odot}$ and radius $R_{eq}=10~\rm{km}$. Models A2 and C2 then
  correspond to stars with rotation period $P=3.07~\rm{ms}$ and
  $P=3.00~\rm{ms}$, respectively. The other stellar parameters are
  $\bar \veps = 0.5$ for the entrainment and $\sigma=0$ for the
  symmetry energy. In the case of model A2 the proton fraction is
  constant, $x_\p =0.1$, while model C2 is stratified with central
  proton fraction $\x_\p(0)=0.1$. The initial configuration
  corresponds to a large glitch with~$\Delta \Omega_\p / \Omega =
  10^{-6}$, as described in the main text.  We run the simulation for
  about $27.25~\rm{ms}$, i.e. about 9 rotation periods, and neglect
  the mutual friction force. From the displayed results, the strong
  effects of the stratification on both the oscillation spectrum and
  gravitational-wave amplitude are evident. \label{fig:gw1}}
\end{figure}

In Fig.~\ref{fig:gw1} we show the characteristic strain $h_c$ for the
A2 and C2 models with $\Delta \Omega_\p / \Omega = 10^{-6}$. The
results refer to a star with mass $M=1.4 M_{\odot}$, radius
$R_{eq}=10~\rm{km}$ and with a low level of mutual friction. We
consider an evolution that lasts for $\sim 27.25~\rm{ms}$ and extract
the signal at $1~\rm{kpc}$\footnote{It is worth noting that we have
  evolved the system for approximately ten rotation periods. We did
  not extend the evolutions because, in reality, one would expect the
  coupling of the two components to start playing a role at this
  stage. If this were not the case and the oscillation prevailed for
  the entire timescale of gravitational-wave damping, then the f-mode
  may last a few seconds. This would be about a factor of 100 longer
  than our evolution, meaning that the effective gravitational-wave
  strain could increase by perhaps an order of magnitude. It would
  still be too weak to be detectable. }. This is roughly the distance
to the Vela pulsar.  The first key result in Fig.~\ref{fig:gw1} is
that the gravitational signal of the C2 model is about \underline{ten
  orders of magnitude larger} than that of the A2 model. This is an
enormous difference, given that these ought to be the same kind of
events. The difference is due to the presence of composition gradients
in the C models. In a stratified model, the co- and counter-moving
degrees of freedom are coupled during the evolution. This coupling is
crucial, since only the co-moving motion generates gravitational
radiation.  The initial data for the pre-glitch lag between neutrons
and protons, in accordance with
equations~(\ref{eq:arel1})-(\ref{eq:arel2}), represent a counter
flow. Hence, in the non-stratified A models these conditions generate
a purely counter-moving motion between the two components that does
not produce any gravitational signal at all.  The fact that the strain
of model A2 is not completely zero in the left panel of
Fig.~\ref{fig:gw1} is likely due to numerical errors. In fact, we have
established that the level of radiation decreases with increased
resolution.  The result shown for model A2 corresponds to an initial
non-corotating configuration where the variation of the total
rotational kinetic energy is $\Delta E_{rot} / E_{rot} \simeq
10^{-20}$. If we increase the precision of the self-consistent field
method that provides the initial data we can lower this value and
consequently the gravitational signal converges to zero. Moreover, the
numerical noise in the simulation excites some oscillation modes that
are related to the co-moving motion. In the left panel of
Fig.~\ref{fig:gw1} we identify the fundamental $l=2$ mode
${}^2\rm{f}$, the first two pressure modes ${}^2\rm{p}_{1}$ and
${}^2\rm{p}_{2}$, and the quasi-radial fundamental mode $\rm{F}$ with
its first overtone $\rm{H}_{1}$.

Let us contrast the results for the non-stratified A2 model with the
results for model C2. In this case, the chemical coupling between the
two fluids introduces a co-moving motion already in the initial
data. The evolutions then generate a larger gravitational strain and
several oscillations modes, like the fundamental $l=0$ and $l=2$ modes
and their respective overtones. In particular, in the right panel of
Fig~\ref{fig:gw1} we note that both ordinary and superfluid modes are
present in the gravitational radiation. This is due to the coupling of
the degrees of freedom (the oscillation modes are no longer purely co-
or counter-moving).  The mode frequencies of the non-rotating model C0
have been compared with the results of~\cite{2002A&A...393..949P}. The
two results agree to better than
1.4\%~\citep[see][]{Passamonti-prep}. We have identified the
oscillation modes of the C2 model by carrying out simulations with
different values for the entrainment $\bar \veps$ and tracking the
superfluid modes as the parameter changes. To this end, we have also
used the analytical formulae determined by~\cite{2009MNRAS.396..951P}.

%-----------------FIG-----------------------------
%----------------------------------------------

%%%%%%%%%%%%%%%%%%%%%%%%%% SEC %%%%%%%%%%%%%%%%%%%%%%%%%%%%%%%%%
\section{Concluding remarks}

We have discussed the dynamics of pulsar glitch events from two,
complementary, points of view. First we constructed a simple model
based on global ``averaging'' of the standard two-fluid equations
including the mutual friction due to superfluid vortices. This
analysis provides a more detailed derivation of the phenomenological
relations that have been used in many discussions of glitches.  In
particular, our final relations clarify how the spin-up time depends
on key parameters like the entrainment. The derivation also highlights
the various assumptions and the restricted validity of the
model. Anyway, for typical values of the parameters (see
Sec.~\ref{sec:4.3}), our model has a glitch rise time shorter than the
upper bound set by current observations.  The model provides a useful
description of the actual glitch event, but it does not account for
the subsequent long-term relaxation (on a timescale of days to months)
of the system.  A key conclusion from our discussion is that the late
stages of evolution requires additional assumptions, most likely,
concerning the repinning of vortices. Understanding this phase better,
e.g. connecting it to the two-fluid hydrodynamics and the averaged
forces that act on the vortices, is an important challenge for the
future.  It seems clear that vortex creep will play a central role
\citep{AI75,1984ApJ...276..325A,1989ApJ...346..823A,1993ApJ...403..285L,1993ApJ...409..345A},
but this mechanism has not yet been discussed in terms of the
macroscopic hydrodynamics.  This issue needs to be addressed if we are
to develop more detailed models of glitch dynamics. We definitely need
to move beyond phenomenology.

As a first step towards hydrodynamic glitch modelling, we have
extended the recent linear perturbation evolution code
of~\citet{2009MNRAS.396..951P} to include the mutual friction and the
perturbed gravitational potential.  Initiated with perturbations that
represent two fluids rotating uniformly at different rates, the
numerical code shows how the system relaxes to co-rotation. We have
analysed this relaxation in detail and demonstrated that the behaviour
is accurately described by the phenomenological model, at least for
non-stratified stellar models. When the star is stratified (e.g. has
varying composition) the relaxation deviates from the simple
model. This is as expected, since the global model was derived under
the assumption of uniform parameters. Of course, the numerical
evolutions provide us with a useful tool for studying the behaviour of
more complex stellar models. In addition, our time evolutions provide
a first insight into the excitation of neutron star oscillations by
glitches. Our results show that a set of axisymmetric modes are
excited by the glitch initial data.  These modes will radiate
gravitational waves\footnote{In principle, the induced oscillations
  may also lead to variations in the electromagnetic signal.  However,
  in order to quantify this effect one would need a more detailed
  analysis of the coupling between the motion of the crust and the
  magnetosphere. Such estimates are beyond the scope of the present
  analysis, but it is worth noting that the oscillation modes that we
  consider are all in the kHz range (much faster than the spin-period)
  meaning that they would not be seen as ``modulations'' of the pulsar
  signal. }, and it is important to establish if the associated
signals may be observable with future detectors. In this respect, our
results are quite pessimistic.  In the cases that we have considered,
the gravitational-wave signal is too weak to be detectable (even with
a third generation of detectors)\footnote{At first sight this
  conclusion seems at variance with the results of
  \citet{2008CQGra..25v5020V}, who consider a different glitch
  scenario. In their (cylindrical) model problem the
  gravitational-waves are associated with the large scale Ekman flow
  that results from a rotational lag between the crust and the core in
  the star. The two mechanisms are obviously different. In particular,
  in our case the event is impulsive and one would expect the signal
  to be burst-like. We certainly cannot envisage the 14 day
  integration suggested by \citet{2008CQGra..25v5020V} to improve the
  signal to noise ratio.  Basically, the model parameters used by
  \citet{2008CQGra..25v5020V} seem rather optimistic.}. However, it is
not clear that this is the final say on the matter. One should keep in
mind that the gravitational-wave strain differs enormously for our two
model configurations. The non-stratified model does not (in principle)
radiate at all, while the stratified model leads to a qualitatively
interesting (albeit weak) signal.  The enormous difference between
these results shows that we need to continue to refine our
modelling. We obviously have to account for the variation of
composition throughout the star, and consider the fact that superfluid
components will only be present in specific density regions. We also
need to understand the nature of the vortex pinning better. In our
models we have assumed that the vortices unpin in a catastrophic
global event. It is far from clear that this is the case in a real
system. It could, for example, be that the unpinning is
localized. This would make the event less symmetric which may enhance
the gravitational-wave signal. We clearly need to understand the
actual mechanism that triggers the glitches better.  The superfluid
instability discussed by \citet{2009PhRvL.102n1101G} is interesting in
this respect, but we need to study this mechanism in more detail to
establish to what extent it can operate in a real neutron star.

To make progress we need to overcome a number of challenges. Yet,
recent developments have provided us with interesting insights and
(most importantly) computational technology that should allow us to
study much more realistic neutron star models in the not too distant
future.

\section*{Acknowledgements}
T.S. is suported by a post-doc fellowship from EU FP6 Transfer of
knowledge project 'Astrophysics of Neutron Stars' (ASTRONS,
MTKD-CT-2006-042722) at Sabanci university.  NA acknowledges support
from STFC via grant number PP/E001025/1.

%%%%%%%%%%%%%%%%%%%%%%%%%%%%%%%%%  BIBLIOGRAPHY  %%%%%%%%%%%%%%%%%%%%%%%%%%%%%%%
\nocite*
%\bibliographystyle{apsrev}
%\bibliographystyle{mn2e}
%\bibliography{references}

%%%%%%%%%%%%%%%%%%%%%%%%%%%%%%%%%  LAST PAGE %%%%%%%%%%%%%%%%%%%%%%%%%%%%%%%
\label{lastpage}

\end{document}